\documentclass[11pt]{article}

\usepackage{amsfonts,amsmath,amssymb,graphics,color}

\newtheorem{lemma}{Lemma}
\newtheorem{defn}{Definition}
\newtheorem{prop}{Proposition}
\newtheorem{thm}{Theorem}

\newtheorem{algm}{Algorithm}

\newcommand{\gap}{\vspace{5mm}}

\newcommand{\C}{\mathbb{C}}
\newcommand{\N}{\mathbb{N}}

\newcommand{\R}{\mathbb{R}}

\newcommand{\I}{\mathrm{i}}
\newcommand{\pr}{\mathbf{P}}

\newcommand{\ffield}{A}
\newcommand{\aitken}{\mathcal{A}}
\newcommand{\catalan}{\mathfrak{c}}
\newcommand{\cumulant}{\mathfrak{k}}
\newcommand{\barrier}{y_+}

\newcommand{\barrierx}{x_+}
\newcommand{\ecoef}{\nu}
\newcommand{\kcoefstuff}{\rho(y,\barrier)}
\newcommand{\fancyh}{h}
\newcommand{\hinfty}{\overline{h}}
\newcommand{\hinftysharp}{\widetilde{h}}
\newcommand{\fancyl}{\ell}
\newcommand{\gothicn}{M}

\newcommand{\gothicz}{\mathfrak{z}}

\newcommand{\riccati}{\mathcal{Q}}

\newcommand{\notthis}[1]{}
\newcommand{\Real}{\mathrm{Re}\,}

\newcommand{\inv}{^{-1}}

\newcommand{\half}{\frac{1}{2}}

\newcommand{\cdl}{\,|\,}

\newcommand{\shalf}{{\textstyle\frac{1}{2}}}

\newcommand{\pderiv}[2]{\frac{\partial{#1}}{\partial{#2}}}
\newcommand{\deriv}[2]{\frac{d{#1}}{d{#2}}}
\newcommand{\dderiv}[2]{\frac{d^2{#1}}{d{#2}^2}}

\newcommand{\pdderiv}[2]{\frac{\partial^2{#1}}{\partial{#2}^2}}

\newcommand{\Beta}{\mathrm{B}}

\newcommand{\bigD}{\mathbf{D}}

\newcommand{\erfc}{\mathrm{erfc}}
\newcommand{\He}{\mathrm{He}}
\newcommand{\sech}{\mathrm{sech}}
\newcommand{\sgn}{\mathrm{sgn}\,}

\begin{document}

\title{\bf Long- and short-time asymptotics of the first-passage time of the Ornstein--Uhlenbeck and other mean-reverting processes}
\author{R. J. Martin\footnote{Department of Mathematics, Imperial College London, South Kensington, London SW7 2AZ, UK}\mbox{ }, M. J. Kearney\footnote{Senate House, University of Surrey, Guildford, Surrey GU2 7XH, UK}\mbox{ } and R. V. Craster$^*$}
\maketitle

\begin{abstract}

The first-passage problem of the Ornstein--Uhlenbeck (OU) process to a boundary is a long-standing problem with no known closed-form solution except in specific cases.
Taking this as a starting-point, and extending to a general mean-reverting process, we investigate the long- and short-time asymptotics using a combination of Hopf-Cole and Laplace transform techniques. As a result we are able to give a single formula that is correct in both limits, as well as being exact in certain special cases. 
We demonstrate the results using a variety of other models.

\end{abstract}

%%%%%%%%%%%%%%%%%%%%%%%%%%%%%%%%%%%%%%%%%%%%%%%%%%%%%%%%%%%%%%%%%%%%%%%%%%%%%%%%%%%%%%%%%%%%%%%%%%%%%%%%%%%%%%%%%%%%%%%%%%%%%%%%%%%%%%%%%%%%%%%%%%%%
%%%%%%%%%%%%%%%%%%%%%%%%%%%%%%%%%%%%%%%%%%%%%%%%%%%%%%%%%%%%%%%%%%%%%%%%%%%%%%%%%%%%%%%%%%%%%%%%%%%%%%%%%%%%%%%%%%%%%%%%%%%%%%%%%%%%%%%%%%%%%%%%%%%%

\section*{Introduction}

The first-passage time of a stochastic process to a boundary is a fundamental problem with applications in queuing theory \cite{iglehart65}, mathematical finance \cite{RISK03}, epidemic models on networks for the spreading of disease and computer viruses  \cite{Lloyd01}, animal or human movement \cite{Gonzalez08}, neuron firing dynamics \cite{Tuckwell05}, diffusion controlled reactions \cite{Szabo80}, controlled kinetics \cite{Benichou10,Godec16}, renewal and non-renewal systems \cite{Ptaszynski18} and much more besides.  Redner \cite{Redner01} presents a physics perspective, and provides a compelling overview of the importance, of many first-passage processes. 

The Ornstein-Uhlenbeck (OU) process \cite{Uhlenbeck30} is the canonical mean-reverting process, with applications in all the above fields; with regard to mathematical finance, it is indispensable in interest-rate modelling \cite{Hull90}.
The boundary problem was studied early on: Darling \& Siegert \cite{Darling53} obtained the Bromwich integral representation for the solution, and made the comment that ``it appears very difficult to invert this transform'' which was prescient given the numerous attempts, with varied success, to invert it analytically.
%, infamously \cite{Scaillet98,Scaillet00}. 
This OU barrier problem, and its mean-reverting generalisations, have therefore been a tantalising, and apparently intractable, long-standing problem that remains of substantial interest: general discussions can be found in \cite{Borodin02,Breiman67,Horowitz85}.
Recent semi-analytic approaches based around an integral equation formulation \cite{Lipton18}, or recursion methods for the moments \cite{Veestraeten15}, indicate the current state-of-the-art.
Here we generate an asymptotic formula valid for long and short times that is also exact in certain cases; in some ways more importantly, we show that the same approximation is not specific to the OU model, but more generally valid under only mild assumptions.

The Feynman--Kac theorem allows us to express the first-passage probability associated with a stochastic process as the solution of a parabolic partial differential equation (PDE) with appropriate initial and boundary conditions. Specifically let
\begin{equation}
dY_t = \kappa \ffield(Y_t) \, dt + \sqrt{2\kappa} \, dW_t,
\label{eq:sde_y}
\end{equation}
be a diffusion, with $\kappa>0$ a constant of dimension $1/$time, %$\inv$, 
and write $\tau =\kappa t$. Then if $F$ is the absorption probability, or equivalently $1-F$ as the survival probability, for a boundary placed at $\barrier$, above the starting-point $y$ (we can always assume $y<\barrier$ without loss of generality), so that
\[
F(\tau,y)=\pr\Big(\max_{0\le t'\le t} Y_{t'} > \barrier\Big), 
\]
then we are to solve
\begin{equation}
\pderiv{F}{\tau} = \ffield(y) \pderiv{F}{y} +  \pdderiv{F}{y},
\label{eq:pde_Fy}
\end{equation}
with initial condition $F(0,\cdot)=0$ and boundary conditions $F(\cdot,\barrier)=1$, $F(\cdot,-\infty)=0$. The p.d.f.\ of the first passage time,  $f=\partial F / \partial \tau$, satisfies the same equation, but with a delta-function initial condition instead.

Of particular interest is the OU process, given by $\ffield(y)=-y$. Despite the fact that the transition density for the unconstrained process is very simple to write down \cite{Cox65}, finding the distribution of its first-passage time to a boundary is much more difficult \cite{Darling53}.
%, and it has been something of a thorn in the side of researchers for the last half-century.
We note that the work of Leblanc, Renault and Scaillet \cite{Scaillet00,Scaillet98}, purporting to give an exact solution to it, is incorrect:
\begin{equation}
f(\tau,y) \stackrel{??}{=} \frac{|y-\barrier| e^{-\tau}}{\sqrt{\pi (1-e^{-2\tau})^3/2}} \exp\left( -\frac{(ye^{-\tau} - \barrier)^2}{2(1-e^{-2\tau})} \right) \qquad \mbox{(Wrong if $\barrier\ne0$)},
\label{eq:wrong}
\end{equation}
though it does correspond with the known result \cite{Pitman81b,Ricciardi88} when the boundary is at equilibrium, a result that can be obtained directly or via the Doob transformation.
The technical reason for the incorrectness of (\ref{eq:wrong}) is that the authors had incorrectly used a spatial homogeneity property of the three-dimensional Bessel bridge process \cite{Yor03}. 
There are more obvious reasons for its incorrectness, which do not require a specialist knowledge of stochastic processes. We  can, of course, simply point out that it is wrong because it fails to satisfy (\ref{eq:pde_Fy})\footnote{The theory was developed without reference to the underlying PDE.}; and, as we shall see from tackling the PDE, in certain limits its behaviour is also wrong. Nonetheless, (\ref{eq:wrong}) is illuminating because in some aspects it is `almost' correct and in others not, a matter on which we expand as our paper unfolds. At the time of writing a simple closed-form solution is not known, and given the close connection between the problem and the parabolic cylinder function, it is likely that there is none. 
This does not preclude the existence of simple \emph{approximate} solutions, or methods that combine analytical techniques with numerical ones. We now turn to these.

The Laplace transform of (\ref{eq:pde_Fy}) is one of the commonest methods for deriving analytical results. However, for all but the simplest stochastic processes it fails to give a neat answer because ultimately a second-order differential equation has to be solved, almost invariably invoking a special function---in the OU case, the parabolic cylinder function---after which an inversion integral has to be carried out. The exponential decay-rate as $\tau\to\infty$ can be obtained from the singularities, provided their position can be identified, and we will devote considerable effort to this in \S\ref{sec:lambda}.
Linetsky \cite{Linetsky04} invokes the Bromwich integral to invert the Laplace transform, using the standard complex analysis technique of collapsing the integral around a series of poles in the left half-plane, but the calculation of their positions and residues is not straightforward. One can develop approximations, but these are likely to be model-specific, so that they are unlikely to be useful for other models; furthermore, in general, the singularities in question may not even be simple poles.
Alili et~al.\ \cite{Alili05} give this same representation and then follow it up with an integral representation that is essentially a Bromwich integral (hence invoking the parabolic cylinder function), and a Bessel bridge representation. 
It is worth noting---recalling the title of our paper---that while the Laplace transform readily conveys information about the long-term asymptotics, it is much less helpful in dealing with the short-term behaviour \cite{Redner01} for which other techniques need to be used.

This brings us on to a second class of techniques, which are time-domain methods that, among other things, establish the short-time behaviour as
\begin{equation}
f(\tau,y) \sim \frac{b}{\sqrt{4\pi\tau^3}} e^{-b^2/4\tau},
\label{eq:levysmirnov}
\end{equation}
where $b=|y-\barrier|$ is the distance to the boundary. This is obvious on  probabilistic grounds, and is most easily derived by applying the Hopf-Cole transformation, \cite{Hopf50}, to (\ref{eq:pde_Fy}) and using dominant balance, as done in \cite{Martin15b} in a different context, and in \S\ref{sec:short} of this paper; alternatively the related WKB approximation can be applied, e.g.\ in \cite{Artime18} where it is explained why (\ref{eq:levysmirnov}) is generic provided the drift term is locally bounded.
Typically, though, these methods do not give information about the long-time behaviour, which as we have said above is exponential. Incidentally Artime et al.\ \cite{Artime18} state in their introduction that there is ``an exponential cut-off  if the domain is bounded'', which is misleading: in this paper we treat semi-infinite domains, and the decay is still exponential.
A recent paper by Lipton \& Kaushansky \cite{Lipton18} uses a transformation to a Volterra integral equation of the second kind, from which (\ref{eq:levysmirnov}) follows by solving Abel's integral equation, and while this does not diagnose the long-term asymptotics it shows how to obtain results simply by solving the Volterra equation numerically.
Indeed, this very appealing paper serves to highlight that the problem is still of both theoretical and practical interest.
Finally, we point out that despite the fact that the first-passage time density is not known in closed form for the OU process, certain aspects of the problem are tractable, such as the moments and/or cumulants (\cite{Veestraeten15}, and later discussion), or certain exponential moments \cite{Ditlevsen07}.

The reader will probably have anticipated that one of our central interests is working out how to combine short- and long-time asymptotics: as is apparent from the above discussion, no existing techniques have yet been successful in doing this.
Our interest is in `global' asymptotics: that is to say, formulae that provide approximations in many different r\'egimes at the same time, rather than having to use one formula for one r\'egime, one for another, and so on.
As one example, Ricciardi \& Sato \cite{Ricciardi88} show that for the OU process, in the far-boundary limit, the first-passage time density is exponential, and Lindenberg et al.~\cite{Lindenberg75} state that this result is generic, which in fact is clear on probabilistic grounds, though as we point out later some technical conditions on $\ffield$ are required. However, it is true only over long time scales, and so we want a single formula that encompasses this and also the short-time behaviour (\ref{eq:levysmirnov}). 
Ideally, we also prefer formulas that make these asymptotic properties immediately  obvious, and in this regard we argue that our final formula (\ref{eq:final}), which is constructed by analysis of many different special cases, satisfies this criterion. Furthermore, the ingredients of this formula are universal, so that although the derivation is guided by the OU model, and is exact in certain cases, it is applicable to a wide class of other diffusions. Accordingly we do not entirely agree with Artime et al.\ \cite{Artime18} that ``general formulations are scarce, since one finds a large variability from one problem to another'': while the number of exactly-solved problems is small our final result, (\ref{eq:final}), possesses a universality that does make it generally applicable.

We should also mention numerical methods, that are very common in, say, financial derivative pricing, and which are used to check the validity of our analytical results. The simplest is to set up a trinomial tree to discretise the process in space and time and then calculate $F$ by forward induction. It is simple to implement and very flexible, in that the boundary can be a different shape, and the dynamics made time-varying; see e.g.\ \cite{Hull_OFOD} for a general introduction. The disadvantage, of course, is that it provides no analytical answers, and the longer the time horizon over which results are needed, the longer it takes.

The paper is organised as follows.
We pursue a general development rather than concentrating on the OU case, even though we make constant reference to it.
Section~\ref{sec:lambda} initially follows the Laplace transform route, but using the Hopf-Cole transformation, i.e.\ the logarithmic derivative in the $y$ direction, it derives a recursion that can be used to find the asymptotic decay-rate $\lambda$ (Theorem~\ref{thm:R3} and Algorithm~1). Along the way it gives formulas for the cumulants (Theorem~\ref{thm:cumul}), to which this analysis is closely related, and give some new results on limiting behaviours in the OU model.
Another consequence is that the far-boundary limit, mentioned above, drops out naturally.
Section~\ref{sec:short} uses the Hopf-Cole transformation again, this time applied to the density itself. This gives a short-time development, which is then extended in such a way as to make the solution behave properly as $\tau\to\infty$, while being exact in certain known special cases. This culminates in the final formula (\ref{eq:final}), which is then demonstrated numerically.
There is, therefore, an important thematic connection between the two halves of the paper, in that both invoke the Riccati equation, in different contexts.
Another link is that the coefficient $\lambda$, expressing the rate of long-term exponential decay, is the subject-matter of the first half, and an important component in the second half.
We complete the paper by suggesting possible further developments.

\section{Long-time asymptotics}
\label{sec:lambda}

We start with some well-known facts and then move on beyond what is common knowledge.

\subsection{Notational preliminaries}

We briefly note at the outset that the general form of the SDE is
\begin{equation}
dX_t = \mu_X(X_t) \, dt + \sigma_X(X_t) \, dW_t;
\label{eq:sde_x}
\end{equation}
by making the substitution 
\[
dy/dx = (2\kappa)^{1/2}\big/\sigma_X(x), \qquad \tau=\kappa t,
\]
also known as the Lamperti transformation, we obtain (\ref{eq:sde_y}). Therefore (\ref{eq:sde_y}) has no less generality and we shall make only occasional reference to (\ref{eq:sde_x}).

By Laplace transforming (\ref{eq:pde_Fy}) in time we get
\[
s\widehat{f}(s,y) - f(0,y) = \ffield(y) \pderiv{\widehat{f}}{y}(s,y) + \pdderiv{\widehat{f}}{y}(s,y);
\]
the asymptotic rate of decay of $f$ is then given by the rightmost singularity of $s \mapsto \widehat{f}(s,y)$.
Write $C_+(s,y)$, $C_-(s,y)$ for the decreasing (in $y$) solutions to the homogeneous problem
\[
\pdderiv{C}{y} + \ffield(y) \, \pderiv{C}{y} = s C(s,y)
\]
that are, respectively, bounded as $y\to-\infty$ and $y\to+\infty$.
If we start below the boundary ($y<\barrier$) then the solution to the Laplace-transformed problem is
\begin{equation}
\widehat{f}(s,y) = \frac{C_+(s,y)}{C_+(s,\barrier)} , \qquad y \le \barrier
\label{eq:LT1}
\end{equation}
while if we start above then we use $C_-$ instead.
The asymptotic rate of exponential decay is then obtained by finding the rightmost singularity of $s \mapsto \widehat{f}(s,y)$. 
Without loss of generality we can concentrate on the case where we start below the boundary.

We define $\psi(y)$ to be the invariant density, so that $\psi'/\psi=\ffield$, and $\Psi$ to be its integral, i.e.\ the cumulative distribution function.
Some examples of interest, which we will analyse in some detail, are:
\begin{itemize}
\item
$A(y)=-y$, OU;
\item
$A(y)=-\sgn y$, dry-friction \cite{Touchette10a};
\item
$A(y)=-\alpha\tanh\gamma y$, giving a sech-power for $\psi$.
\end{itemize}
The last example reduces, in opposite extremes, to the first two.

\subsection{Comment on the OU case}

We briefly discuss the OU case, as this helps the reader link our work to previous literature e.g.~\cite{Ricciardi88}, and we note that \cite{Abramowitz64} contains background details of the parabolic cylinder, Tricomi, Kummer functions that are used.
We have
\[
C_\pm(s,y) = \bigD_s (\pm y),
\]
where $\bigD_s(y)$, a relative of the parabolic cylinder function, is variously defined as follows. First, as an integral transform,
\begin{equation}
\bigD_s(y) = \frac{1}{\Gamma(s)} \int_0^\infty u^{s-1} e^{yu-u^2/2} \, du, \qquad \Real s>0,
\label{eq:LT1.ou}
\end{equation}
and by analytic continuation elsewhere, for example by means of the recursion (immediate from the above)
\begin{equation}
\bigD_s(y) = -y \bigD_{s+1}(y) + (s+1)\bigD_{s+2}(y).
\label{eq:pcfrecur}
\end{equation}
(N.B. The definition is convenient, but nonstandard.)
Alternative representations use the Kummer function $M$ (also commonly known as $\Phi$ or $\mbox{}_1F_1$),
\[
\bigD_s(y) = \frac{2^{s/2-1} \Gamma(\frac{s}{2})}{\Gamma(s)}  M { \textstyle (\frac{s}{2},\half,y^2/2) } + \frac{2^{(s-1)/2} \Gamma(\frac{s+1}{2})}{\Gamma(s)}  y \cdot { \textstyle M (\frac{s+1}{2},\frac{3}{2},y^2/2) } 
\]
or, in terms of the Tricomi function
\[
U(a,b,z) = \frac{1}{\Gamma(a)} \int_0^\infty  e^{-zt} t^{a-1} (1+t)^{b-a-1} \, dt, \qquad 0 \le \arg z < \pi/2
\]
we have
\[
\bigD_s(-y) = 2^{-s/2} \textstyle 
U(\frac{s}{2}, \half, y^2/2), \qquad 0 \le \arg y < \pi
\]
where the principal branch of $U(\cdot,\cdot,w)$ is on the cut plane $\{0\le\arg w < 2\pi\}$, i.e.\ the cut is just below the positive real axis\footnote{It is then defined by analytic continuation elsewhere; note that the RHS does not have a branch-point at $y=0$, but because $U$ does, the RHS defines a function on two disconnected copies of $\C$.}.
More usually it is written in terms of the parabolic cylinder function $D_s$ as
\[
\bigD_s(y) = e^{y^2/4} D_{-s}(-y).
\]
As consequences, we note\footnote{$\phi,\Phi$ denote the density and cumulative of the standard Normal distribution; $\He_r$ is the $r$th Hermite polynomial; $\N$ denotes the set of positive integers, and $\N_0$ the same with 0 included.}
\[
\bigD_1(y)=\Phi(y)/\phi(y); \qquad
\bigD_s'(y) = s \bigD_{s+1}(y) ; \qquad
\bigD_{-r}(y)=(-)^r\He_r(y), \quad r\in \N_0 .
\]
These allow the results of Ditlevsen \cite{Ditlevsen07} to be derived.
Another useful result, related to the idea of analytic continuation, is the reflection formula, analogous to that of the Gamma function and derived in the same way, but not nearly as well-known, to the extent that it appears to be a new result (see Appendix) despite integral representations of products of parabolic cylinder functions remaining of interest in the special functions community \cite{Nasri15,Veestraeten17b,Veestraeten17a}:
\begin{eqnarray}
\bigD_s(y) \bigD_{1-s}(y) &=& \sum_{k=0}^\infty  \frac{\Gamma(k+s)\Gamma(k+1-s)}{k! \Gamma(s)\Gamma(1-s)}  \bigD_{2k+1}(y)
\label{eq:pcfrecur2} \\
&=& \int_0^\infty  {}_2F_2 (s, 1-s; 1,\shalf; z^2/4) \, e^{yz-z^2/2} \, dz.
\label{eq:pcfrecur3}
\end{eqnarray}
Obviously (\ref{eq:pcfrecur2}) is another way of extending $\bigD_s(y)$ to the left half-plane (of $s$), doing it `in one go' rather than recursively in the way that (\ref{eq:pcfrecur}) does.

\subsection{Definition of $\lambda$, and some technical conditions}

We now return to the general case.
The rate of exponential decay in the long-time limit is obtained by finding the position of the rightmost singularity of $\widehat{f}$; more formally we define
\begin{equation}
\lambda = \sup \{ a : \widehat{f} \mbox{ analytic for } \Real s>-a \} \ge 0,
\end{equation}
which in principle depends on the starting-point and the boundary, as well as on $\ffield$. In the OU case the rightmost singularity is caused by a zero of $s\mapsto\bigD_s(\barrier)$, which causes $\widehat{f}$, see Eq.~\eqref{eq:LT1}, to have a simple pole. In general, the singularity might not be a pole, but some important general results can be formulated.

We define $R_{s=a}(f)$ to be the radius of differentiability of an analytic function $f$ at the point $s=a$, i.e.\ the radius of convergence of its Taylor series in $s$.
The following shows that rather than finding the strip of analyticity, we can use the radius about the origin:
\begin{lemma}
\label{lem:R1}
$\lambda = R_{s=0}\big[ \widehat{f} \big]$: the asymptotic rate of exponential decay of the first-passage time density is the radius of differentiability of $\widehat{f}(s)$ at the origin.
\end{lemma}
\noindent
Proof. This is elementary and it is sufficient to prove that the closest singularity to the origin is on the negative real axis, rather than a pair of complex-conjugates at $-\lambda \pm \I\omega$, which would make $R_{s=0}[\widehat{f}]=\sqrt{\lambda^2+\omega^2}>\lambda$. But $\widehat{f}$ is the Laplace transform of a probability density $p$ say, and hence of a nonnegative function, so this is impossible. For,
\[
\big| \widehat{f}(-\lambda+\I\omega,z) \big| \le \int_0^\infty e^{\lambda \tau} p(\tau) \, d\tau = \widehat{f}(-\lambda);
\]
so if it is analytic at $s=-\lambda$ then it is analytic at $s=-\lambda\pm\I\omega$. $\Box$

\gap

\begin{defn}
The first-passage problem is said to be completely absorbing, if from any starting-point the process almost surely hits the boundary eventually. It is completely absorbing iff $\widehat{f}(0,y)=1$ for all $y$.
\end{defn}

\begin{lemma}
\label{lem:CA}
The first-passage problem is completely absorbing if the following two conditions are met:
\begin{itemize}
\item
 $\liminf_{y\to-\infty} \ffield(y)\ge0$;
\item 
$\inf_{y<\barrier} \ffield(y)>-\infty$.
\end{itemize}
\end{lemma}
\noindent
Proof.
The easiest way to see this is via a probabilistic argument: the first condition shows that the process is recurrent, and the second shows that its probability of hitting the boundary is positive. Therefore the survival probability must decay to zero.
$\Box$

\gap
\noindent
A slightly stronger condition will be needed shortly:
\begin{defn}
We write $\ffield\in \mathfrak{S}_-$ if $\lim_{y\to-\infty} -y \ffield(y)=+\infty$, and 
$\ffield\in \mathfrak{S}_+$ if the same limit holds as $y\to+\infty$.
%In either case we say that the reversion is semi-strong.
If additionally $\ffield'(y)/\ffield(y)\to0$ and $\ffield'(y)/\ffield(y)^2\to0$ as $y\to-\infty$ then we write $\ffield\in\mathfrak{S}_-^*$; similarly for $y\to+\infty$. 
\end{defn}
The first condition ensures that the reversion speed does not decay too rapidly at $\pm\infty$. Models such as the OU and dry-friction are in $\mathfrak{S}_\pm$ but $-y/(1+y^2)$, which was considered in \cite{Martin15b}, is not. The condition implies that the invariant density, $\psi(y)$, decays faster than any power of $y$, and in most cases it decays exponentially, but it may not do so, for example when $\ffield(y)=-\ln(1+y^2)/y$.
The second condition ($\ffield'(y)/\ffield(y)\to0$ and $\ffield'(y)/\ffield(y)^2\to0$) will be needed later when we need to bound the variation of $\ffield$, essentially because we shall need to approximate the integral of $\ffield(z)$, for $z$ up to some value $y$, with an expression depending only on  $\ffield(y)$.
Informally, most `sensible' force-fields obey this extra condition\footnote{A function that is in $\mathfrak{S}_+$ but not in $\mathfrak{S}_+^*$ is this sawtooth: $\ffield(y)=-1$ for $2n\le y < 2n+1$ and $\ffield(y)=-2$ for $2n+1 \le y < 2n+2$, $n\in\N$.}.

\notthis{
\gap
The next condition is a minor one designed to remove pathological cases, and ensures that the force-field $\ffield$ behaves sensibly at $\pm\infty$:
\begin{defn}
\label{defn:RC}
We say that $\ffield$ is regular at $y=\pm\infty$ if $\ffield'(y)/\ffield(y)^2\to0$ in that limit.
\end{defn}
}% end

\begin{lemma}
\label{lem:RC}
If $\ffield\in\mathfrak{S}_+^*$ then for $r\in\N$,
\[
\int_x^y  \frac{dz}{\ffield(z)^{r-1} \psi(z)^r} \sim \frac{-1}{r \ffield(y)^r \psi(y)^r}, \qquad y\to+\infty,
\]
for any $x$ for which the integral is defined\footnote{Essentially we need to make sure that $\ffield(z)$ is not zero in the range of integration. By hypothesis this will hold for $z$ sufficiently large.}.
\end{lemma}
\noindent
Proof. We have
\[
\int_x^y  \frac{dz}{\ffield(z)^{r-1} \psi(z)^r} =
-\frac{1}{r} \int_x^y \frac{1}{\ffield(z)^r} \deriv{}{z} \left( \frac{1}{\psi(z)^r} \right)  dz
= \left[ \frac{-1}{r\ffield(z)^r \psi(z)^r} \right]_x^y 
- \int_x^y  \frac{\ffield'(z)/\ffield(z)^2}{\ffield(z)^{r-1} \psi(z)^r} \, dz 
\]
and because $\ffield'(y)/\ffield(y)^2\to0$, the second term on the RHS is small compared with the LHS, which is what we are trying to approximate.
$\Box$

\subsection{The logarithmic derivative and its expansion}

The logarithmic derivative of the function $C_+(s,z)$ is going to be central to the theory:
\begin{equation}
H(s,z) = - \pderiv{}{z} \log C_+(s,z),
\end{equation}
and the following result gives some of its more important properties.
\begin{prop}
\label{prop:1}
The following applies in general.
\begin{itemize}
\item[(i)]
Let $f'(t,z)$ denote the $z$-derivative of the first-passage time density at the boundary (where the boundary is placed at $z$). Then for $r\in\N$ the Laplace transform of $t\mapsto-t^rf'(t,z)$ is $(-\partial/\partial s)^rH(s,z)$, which is $>0$.
So $H$ is analytic for $\Real s >$ some $\widehat{s}\le0$ and singular at $s=\widehat{s}$.
\item[(ii)]
$H$ satisfies the Riccati equation
\begin{equation}
\pderiv{H}{z} - H^2 + AH = -s .
\label{eq:ricc1}
\end{equation}
\item[(iii)]
In the formal expansion
\begin{equation}
H(s,y) = \sum_{r=0}^\infty (-s)^r \fancyh_r(y),
\label{eq:Hexpansion}
\end{equation}
we have
\[
\fancyh_0(y)= - \pderiv{}{y} \ln p_\infty(y;\barrier) \le 0
\]
where $p_\infty(y;\barrier)$ denotes the probability that the boundary is eventually hit, starting from $y<\barrier$; note that the above expression does not depend on $\barrier$. Also $\fancyh_r>0$ for $r\ge1$.
\item[(iv)]
We have
\begin{equation}
\fancyh_1(y) = \frac{1}{\psi(y)p_\infty(y)^2} \int_{-\infty}^y \psi(z)p_\infty(z)^2 \, dz ,
\label{eq:recurh1}
\end{equation}
which equals $\Psi(y)/\psi(y)$ in the completely-absorbing case, and
\[
\fancyh_r'(y) = -A(y) \fancyh_r(y) + \sum_{k=0}^{r} \fancyh_k(y)\fancyh_{r-k}(y), \qquad r \ge2.
\]
If  $\ffield\in\mathfrak{S}_-$, we have
\[
\fancyh_1(y) = o(|y|), \qquad y\to-\infty,
\]
and so, as the problem is completely absorbing in that case,
\begin{equation}
\fancyh_r(y) = \frac{1}{\psi(y)} \int_{-\infty}^y \sum_{k=1}^{r-1} \fancyh_k(z)\fancyh_{r-k}(z) \, \psi(z)  \, dz, \qquad r \ge2,
\label{eq:recurh2}
\end{equation}
so that
\begin{equation}
\fancyh_r(y) \sim \catalan_{r-1} \fancyh_1(y)^{2r-1}, \qquad y\to-\infty
\end{equation}
where $\catalan_r=\frac{(2r)!}{r!(r+1)!}$ is the $r$th Catalan number.

\noindent
If additionally $\ffield\in\mathfrak{S}_+^*$ then
\begin{equation}
\fancyh_r(y) \sim \big({-\ffield(y)}\big)^{1-r} \psi(y)^{-r} , \qquad y\to+\infty.
\label{eq:fancyh_infty}
\end{equation}

\item[(v)]
$R_{s=0}[H(s,z)]$ is monotone decreasing in $z$.

\item[(vi)]
$R_{s=0}[H(s,z)] = \lim_{r\to\infty} \fancyh_r(z)/\fancyh_{r+1}(z)$: the ratio of successive terms gives the radius of convergence (and hence $\widehat{s}$ as defined in part (i)).

\item[(vii)] The large-$s$ behaviour is
\begin{equation}
H(s,z) \sim -\sqrt{s}, \qquad s\to+\infty.
\label{eq:Has}
\end{equation}

\end{itemize}
\end{prop}

\noindent
Proof. Part (i) is straightforward (note that $f'<0$; the inequality is untrue for $r=0$ because $f'$ is singular of order $\tau^{-3/2}$ as $\tau\to0$, so does not have a Laplace transform). Part (ii) is immediate from the backward equation, while (iii) is immediate from (i) and (\ref{eq:LT1}).

The first part of (iv) follows from (ii). 
From
\[
\Psi(y) = \int_{-\infty}^y \psi(z)\, dz = y\psi(y)
 - \int_{-\infty}^y z\ffield(z) \psi(z)\, dz ,
\]
the following holds when $\ffield\in\mathfrak{S}_-$: let $c>1$, so as $y\to-\infty$ we have
$\Psi(y) > y\psi(y) +  c \Psi(y)$, and so $\fancyh_1(y)<-y/(c-1)$, which proves that $\fancyh_1(y)=o(|y|)$.
This condition ensures that the integral (\ref{eq:recurh2}) converges, as then $\psi(y)$ decays faster than any power of $|y|$.
The asserted asymptotic behaviour of $\fancyh_r$ as $y\to-\infty$ follows by induction: it is trivial when $r=1$, so let $r\ge2$ and suppose that it holds for all lower values of $r$.
Then
\begin{eqnarray*}
\fancyh_r(y) &\sim& \frac{1}{\psi(y)} \int_{-\infty}^y
\sum_{k=1}^{r-1} \catalan_{k-1} \catalan_{r-k-1} \fancyh_1(z)^{2r-2} \psi(z) \, dz \\
&=& \frac{\catalan_{r-1}}{\psi(y)} \int_{0}^{\Psi(y)} \fancyh_1\big( \Psi\inv(u) \big)^{2r-2} \, du \\
&\sim&  \catalan_{r-1} \frac{\Psi(y)}{\psi(y)} \fancyh_1(y)^{2r-2}
\end{eqnarray*}
as required.
When additionally $\ffield\in\mathfrak{S}_+^*$ we have as $y\to+\infty$
\begin{eqnarray*}
\fancyh_r(y) &\sim& \frac{1}{\psi(y)} \int_{-\infty}^y \sum_{k=1}^{r-1} \fancyh_k(z) \fancyh_{r-k}(z) \psi(z)^{1-r} \, dz \\
&\sim& \frac{r-1}{\psi(y)} \int_{x}^y \big( {-\ffield(z)} \big)^{2-r} \psi(z)^{1-r} \, dz
\end{eqnarray*}
for any $x$ satisfying $z>x\Rightarrow\ffield(z)<0$, and the result then follows from Lemma~\ref{lem:RC}.

As to part (v), we note that by usual arguments on analytic functions,
\[
R_{s=0}[H(s,z)] =  \liminf_{r\to\infty} |\fancyh_r(z)|^{-1/r} .
\]
Now write $\fancyl_r=\fancyh_r^{1/r}$, so that
\[
\fancyl_r'(z) = \frac{-\ffield(z) \fancyl_r(z)}{r} + \frac{\fancyl_r(z)}{r \fancyh_r(z)} \sum_{k=1}^{r-1} \fancyh_k(z) \fancyh_{r-k}(z);
\]
\notthis{
\[
\fancyl_r' = \frac{1}{r} \left[ -\ffield \fancyl_r + \fancyl_r^{1-r} \sum_{k=1}^{r} \fancyl_k^k \fancyl_{r-k}^{r-k} \right] .
\]
Using the AM-GM inequality,
\[
\fancyl_r' \ge \frac{\fancyl_r}{r} \left[ -\ffield + (r-1) \fancyl_r^{-r} \big( \fancyl_1^1 \fancyl_2^2 \cdots \fancyl_{r-1}^{r-1} \big)^{2/(r-1)} \right];
\]
} % end
the first term tends to zero as $r\to\infty$ and the second is positive. As the radius of convergence is $\liminf_{r\to\infty} 1/\fancyl_r(z)$, the result is proven.

Part (vi) follows from using Cauchy's integral formula and observing that the dominant contribution comes from near the singularity.

Part (vii) follows by dominant balance, as $H^2$ equates to $s$ in the limit. (The negative root needs to be taken as otherwise $H$ is non-decreasing.) This implies behaviour near the boundary of
\[
-f'(\tau,z) \sim \frac{1}{\sqrt{4\pi \tau^3}}, \qquad \tau\to0
\]
and is a consequence of the diffusive behaviour: so it works for any model, regardless of the drift or boundary position.
$\Box$

As an aside, notice that in the OU case with the boundary at equilibrium,
\begin{equation}
H(s,0) = -\sqrt{2}\, \frac{\Gamma\big(\frac{s+1}{2})}{\Gamma\big(\frac{s}{2}\big)}
\end{equation}
which is seen to have the advertised behaviour.

\gap

The centrality of the function $H(s,z)$ is contained in the following result, which shows that $\lambda$ is simply the radius of convergence of $H(s,\barrier)$:

\begin{lemma}
\label{lem:R2}
$R_{s=0}\big[ \widehat{f} \big] = 
R_{s=0} \big[ H(s,\barrier) \big]$.
\end{lemma}

\noindent
Comment.
Thus $\lambda$, the asymptotic rate of exponential decay of the first-passage time density, depends on the position of the boundary, and not on the starting-point, which is  unsurprising as, over time, the process forgets about where it started. 
The fact that $R_{s=0}\big[ H(s,\barrier) \big]$ is monotone decreasing in $\barrier$ now comes as no surprise: if the boundary is brought closer, it must be hit at least as rapidly, so $\lambda$ must increase (or perhaps stay the same).
In practical terms this result is important because we can obtain $\lambda$ from the sequence $\big(\fancyh_r(\barrier)/\fancyh_{r+1}(\barrier)\big)$.

In the particular case of OU, Elbert and Muldoon \cite{Elbert08} showed that the position of the rightmost zero of $s\mapsto \bigD_s(z)$ is monotonic in $z$, by direct analysis of that function using results known as Nicholson integrals \cite{Nasri15,Veestraeten17a,Veestraeten17b}. The above development adds to their work by showing that it is a direct consequence of probability theory and also that the Riccati equation provides another way of analysing the problem. What is perhaps remarkable is that we can make inferences about the leading zero of $s\mapsto \bigD_s(z)$ even if we have no idea of how to calculate that function.
As a consequence the Riccati equation gives generic results, i.e.\ even when $\ffield$ is not linear.

\gap

\noindent
Proof of Lemma~\ref{lem:R2}.
We have
\[
\widehat{f}(s,y;\barrier) = \exp \int_y^{\barrier} H(s,z) \, dz.
\]
If $H$ is analytic then so is the integral on the RHS, so we have proved `$\ge$' in the assertion: $\widehat{f}$ is at least as differentiable as $H(s,z)$.
Conversely, if $\widehat{f}$ is analytic for $s>\widehat{s}$ then it is real and positive, so a continuous branch of $\log \widehat{f}$ exists and then $s\mapsto H(s,\barrier)$ is also analytic.
$\Box$

\gap

Accordingly, we have:
\begin{thm}
\label{thm:R3}
The asymptotic exponential decay-rate is given by
\[
\lambda = \lim_{r\to\infty} \fancyh_r(\barrier)/\fancyh_{r+1}(\barrier).
\]
If $\ffield\in\mathfrak{S}_-$ then
\begin{equation}
\lambda\sim \frac{\psi(\barrier)^2}{4\Psi(\barrier)^2},  \qquad \barrier \to-\infty.
\label{eq:lambda-}
\end{equation}
If additionally $\ffield\in\mathfrak{S}_+^*$ then\footnote{Recall $\ffield=\psi'/\psi$.}
\begin{equation}
\lambda \sim - \psi'(\barrier), \qquad \barrier\to+\infty .
\label{eq:lambda_asymp}
\end{equation}
Referring to the dimensional form\footnote{So that, for (\ref{eq:lambda_asymp2}) alone, $\lambda$ refers to time $t$ rather than $\tau$: the density decays as $e^{-\lambda t}$.} (\ref{eq:sde_x}), these are
\begin{equation}
\lambda\sim \frac{\sigma_X(\barrierx)^2 \psi_X(\barrierx)^2 }{8\Psi_X(\barrierx)^2}, 
\qquad
\lambda\sim -\mu_X(\barrierx) \psi_X(\barrierx) .
\label{eq:lambda_asymp2}
\end{equation}
\end{thm}
\noindent
Proof. Immediate from Lemmas~\ref{lem:R1},\ref{lem:R2} and Prop.~\ref{prop:1}(iv,vi); note that $\lim_{r\to\infty} \catalan_r^{1/r}=4$. 
$\Box$

\subsection{Far-boundary limit $\barrier\to+\infty$}

Ricciardi \& Sato \cite{Ricciardi88}, in their work on the OU process, show that as $\barrier\to+\infty$ the distribution of the first passage time is asymptotically exponential. More precisely, the density of $\tilde{\tau}=\lambda\tau$, for fixed $\tilde{\tau}$, tends to $\exp(-\tilde{\tau})$. For convenience we shall call this the rescaling limit. They prove this by analysing the Taylor series of the characteristic function, and also empirically compute moments, finding agreement. Lindenberg et al.~\cite[eq.(74)]{Lindenberg75} suggest that this result is generic, by means of an eigenfunction expansion.

Before going into the technical details it is worth making an intuitive argument, as follows.
As the boundary is pushed further away, and we look at the process on progressively longer time scales (of order $1/\lambda$), the process becomes less autocorrelated. We are therefore observing the waiting-time for the first occurrence of a Poisson process, and that must be exponentially distributed.
That $\lambda \sim -\psi'(\barrier)$ is also intuitively sensible: the relative rate of change of the survival probability must be closely linked to the probability density of the unconstrained process in the vicinity of the boundary.

%This principle, incidentally, justifies the effort we have devoted to finding $\lambda$, as in the rescaling limit it is the only thing that matters.

\notthis{

In terms of what we have done above, we observe that
\begin{eqnarray*}
\pderiv{}{z}  \log \widehat{f}(s) = H(s,\barrier) 
&=& \sum_{r=1}^\infty (-s)^r \fancyh_r(y) \\
&\sim&  -\ffield(y) \sum_{r=1}^\infty \left( \frac{s}{\ffield(y)\psi(y)} \right)^r 
 \mbox{ (by Prop.~\ref{prop:1}(iv))} \\
&=& \frac{-s/\psi(\barrier)}{1-s/\psi'(\barrier)}
\end{eqnarray*}
where we recall that $\ffield=\psi'/\psi$.
By the regularity condition (Defn.~\ref{defn:RC}) we have
\[
\frac{\psi''(\barrier)\psi(\barrier)}{\psi'(\barrier)^2} \to 1
\]

} % end

We can, in fact, use Ricciardi \& Sato's working to derive a general result. Indeed, let $\{\cdot\}_r$ denote the coefficient of $s^r$ in the Taylor series (around $s=0$). Then, with the boundary placed at $z$ and the starting-point at $y$ we have by (\ref{eq:fancyh_infty}), 
\[
\left\{  \pderiv{}{z} \log \widehat{f}(s,y;z) \right\}_r  = (-)^r\fancyh_r(z) \sim \frac{-\ffield(z)}{\psi'(z)^r}, \qquad z\to+\infty,
\]
for each $r\in\N$.
If $\ffield\in\mathfrak{S}_+^*$ then $\ffield'(z)/\ffield(z) \to 0$, so we can replace $\ffield=\psi'/\psi$ with $\psi''/\psi'$ and the above equation becomes (cf.~\cite[eq.(37)]{Lindenberg75})
\[
\left\{ \pderiv{}{z} \log \widehat{f}(s,y;z) \right\}_r \sim \frac{-\psi''(z)}{\psi'(z)^{r+1}}, \qquad z\to+\infty.
\]
Integrating both sides from $z=y$ to $\barrier$ gives
\[
\left\{ \log \widehat{f}(s,y;\barrier) \right\}_r  \sim  \frac{1}{r\psi'(\barrier)^r} + \beta_r, \qquad \barrier\to+\infty,
\]
where $\beta_r$, the `constant of integration', depends on $y$ but not on $\barrier$. Accordingly
\[
\log \widehat{f}(s,y;\barrier)  \sim  - \log \big(1-s/\psi'(\barrier)\big) + B_y(s)
\]
where $B_y(s)=\sum_{r=1}^\infty \beta_r s^r$; as $\widehat{f}(0)$ is necessarily zero, we must have $B_y(0)=0$, so $\beta_0=0$.
Write $s=-\psi'(\barrier)\tilde{s}$ with $\tilde{s}$ fixed, and let $\barrier\to+\infty$. Then the $B_y(s)$ term disappears, and $\widehat{f}(s)$ is recognised as the Laplace transform of the exponential distribution of mean $1/\lambda=-1/\psi'(\barrier)$, which is what we were to show.

Notice that we have assumed, in our derivation, that $\ffield\in\mathfrak{S}_+^*$. A simple example of a force-field that does not obey this condition is the arithmetic Brownian motion, $\ffield(y)=\mu>0$. In this case, the first-passage time density (using $(t,x)$ coordinates) is the inverse Gaussian distribution \cite{Seshadri93}, 
\begin{equation}
f(t,x) =  \frac{\barrierx-x}{\sqrt{2\pi \sigma^2 t^3}} \exp \left( \frac{-(\barrierx-x-\mu t)^2}{2\sigma^2 t} \right) ;
\label{eq:invgauss}
\end{equation}
the rate of exponential decay is $\mu^2/2\sigma^2$ which does not tend to zero as $\barrierx\to+\infty$ and the rescaling limit does not apply. Indeed, it is clear in this case that any kind of rescaling will preserve the front factor of $t^{-3/2}$, and there is no way of ending up with an exponential distribution.
In processes that do not mean-revert, the above argument on autocorrelation fails, and with it the assertion that an exponential distribution will necessarily arise in the rescaling limit.
This problem was not addressed in \cite[\S4.4.1]{Lindenberg75} who have, in effect, claimed too great a degree of generality in their proof.

In fact, this observation has pointed out a continuing difficulty over the $\tau^{-3/2}$ factor.
As intimated in the Introduction, we should be expecting the behaviour shown in (\ref{eq:levysmirnov}) for \emph{short} time; so how can this be reconciled with the rescaling limit that, when $\ffield\in\mathfrak{S}_+^*$, causes the $\tau^{-3/2}$ behaviour to vanish, but when $\ffield\notin\mathfrak{S}_+^*$, may let it remain?
We are asking for an approximation to the first-passage time density that works on time scales  $\tau\sim1/\lambda$ and also $\tau=O(1)$, and this ansatz does the trick:
\begin{equation}
f(\tau,y) \propto \frac{e^{-b^2\theta\sqrt{q}/2(1-q)}}{(1-q)^{3/2}} \cdot e^{-\lambda\tau}, \qquad q=e^{-2\theta \tau},
\label{eq:rs_ansatz}
\end{equation}
where $\theta$ sets the rate of mean reversion.
It is clear that the behaviour is the same as (\ref{eq:levysmirnov}) for short time ($\theta\tau\ll1$).
In the rescaling limit, the first part of this expression disappears, provided $\theta>0$, as in effect $q$ is replaced by zero, and we end up with $e^{-\tilde{\tau}}$.
But if we first let $\theta\to0$, as happens in the arithmetic Brownian motion, we instead obtain the inverse Gaussian distribution, and the $\tau^{-3/2}$ factor persists in any scaling limit.
As it happens, we do end up with something like (\ref{eq:rs_ansatz}), and it will naturally emerge from the work in \S\ref{sec:short}.

\subsection{Moments and cumulants of the first-passage time density}

The cumulants ($\cumulant_r$) of the first-passage time\footnote{We are using `dimensionless time' ($\tau$) here. Obviously those for dimensional time ($t$) are $\cumulant_r/\kappa^r$.} relate directly to the $(\fancyh_r)$, by
\begin{equation}
\frac{\cumulant_r}{r!} = \int_{y}^{\barrier} \fancyh_r(z) \, dz.
\end{equation}
Hence:
\begin{thm}
\label{thm:cumul}
All the cumulants of the first-passage time are positive and are obtained from the recurrence (\ref{eq:recurh2}); the mean and variance are 
\begin{equation}
\int_{y}^{\barrier} \frac{\Psi(z)}{\psi(z)} \, dz, \qquad 
\int_{y}^{\barrier} \frac{2}{\psi(z)} \int_{-\infty}^z \frac{\Psi(w)^2}{\psi(w)} \, dw \, dz
\end{equation}
respectively. If $\ffield\in\mathfrak{S}_- \cap \mathfrak{S}_+^*$ then in the far-boundary limit $\cumulant_r\sim\lambda^{-r}$.
$\Box$
\end{thm}

In the OU case we note that $\fancyh_1$ admits the integral representation 
\[
\fancyh_1(z) = \int_0^\infty e^{-u^2/2} e^{zu} \, du
\]
from which the mean is expressible in two ways,
\begin{equation}
\cumulant_1 = \int_y^{\barrier} \frac{\Phi(z)}{\phi(z)} \, dz = \int_0^\infty e^{-u^2/2} \left( \frac{e^{\barrier u} - e^{yu}}{u} \right) \, du.
\label{eq:ouk1}
\end{equation}
We briefly discuss how this behaves in different r\'egimes.
One is what, in dimensional coordinates, would be called the low-reversion r\'egime ($\kappa$ `small'), and in dimensionless coordinates is obtained by making $|y|$ and $\barrier$ small.
By expanding $\Phi,\phi$ around the origin we have
\[
\cumulant_r \sim \left(\sqrt{\frac{\pi}{2}} + \frac{\barrier+y}{2} \right) (\barrier-y).
\]
Another is what, in dimensional coordinates, would be called the low-volatility r\'egime ($\sigma$ `small'), and in dimensionless coordinates is obtained by making  $|y|$ and/or $|\barrier|$ large. This subdivides into three cases. 
First, the case that is often described as `sub-threshold' is $\barrier\gg1$ and we have already given the asymptotic as 
\[
\cumulant_1 \sim \frac{(2\pi)^{1/2}  e^{\barrier^2/2}}{\barrier}
\]
regardless of the starting-point $y$.
Next, in the `supra-threshold' case, in which the boundary lies between the starting-point and equilibrium, we have $y,\barrier\to-\infty$ and then
\[
\cumulant_1 \sim \frac{1}{2} \ln \frac{y^2}{\barrier^2} +  \sum_{r=1}^\infty \frac{(-)^r(2r-1)!!}{2r\, z^{2r}} (\barrier^{-2r} - y^{-2r}) ;
\]
the first term is recognisable as the time taken to hit the boundary if the volatility were zero, i.e.\  $Y_t=ye^{-\kappa t}$, and the leading-order correction ($r=1$ term) is negative, suggesting that the presence of volatility causes the boundary to be hit earlier than that, on average. 
Finally the medial case is when the boundary is at equilibrium ($\barrier=0$): as $y\to-\infty$, we have
\[
\cumulant_1 \sim  \frac{\ln (2y^2) + \overline{\gamma}}{2} - \sum_{r=1}^\infty \frac{(-)^r(2r-1)!!}{2r\, y^{2r}} 
\]
with $\overline{\gamma}$ denoting Euler's constant.
This is obtained by substituting $-u/y$ for $u$ in the second integral in (\ref{eq:ouk1}) and invoking Plancherel's identity\footnote{In effect we have a continuous version of the Poisson summation formula.}:
\[
\int_{-\infty}^\infty e^{-u^2/2y^2} \left( \frac{1-e^{-u}}{u} \mathbf{1}_{u>0} \right) du
=
\frac{|y|}{\sqrt{2\pi}} \int_{-\infty}^\infty e^{-y^2\omega^2/2} \log \left(\frac{1+\I\omega}{\I\omega}\right) \, d\omega .
\]
%where on the RHS the branch-cuts in the $\log()$ term are taken so as to make the integrand regular in the lower half-plane (and the contour runs just below the real axis).
Then the $\log(\I\omega)$ term generates
\[
\frac{-|y|}{\sqrt{2\pi}} \int_{-\infty}^\infty e^{-y^2\omega^2/2} \ln |\omega| \, d\omega =  \frac{ \ln (2y^2) + \overline{\gamma}}{2}
\]
and the $\log(1+\I\omega)$ term, upon expansion in a Taylor series around $\omega=0$, delivers the rest (and is clearly related to the asymptotic expansion of the error function).

%For the proof, recall $d\mathfrak{k}_1/dy=-\Phi(y)/\phi(y)$ and integrate the asymptotic expansion of that function; this shows that the LHS and RHS differ by a constant.

\subsection{Algorithm}

We now turn to matters of numerical computation.
The recursion (\ref{eq:recurh1},\ref{eq:recurh2}) makes the numerical calculation of the $\fancyh_r(z)$ easy, regardless of the choice of force-field $\ffield$. In more detail, starting from $r=2$, we have:

\begin{algm}
Evaluation of the Taylor series (\ref{eq:Hexpansion}) of $H(s,z)=(-\partial/\partial z)\log C_+(s,z)$, when $\ffield\in\mathfrak{S}_-$. First note that $\fancyh_1$ is  given by (\ref{eq:recurh1}). Set $r=2$. Then:
\begin{itemize}
\item[(i)]
Set $z$ equal to large negative value $Z$ and approximate the integral in (\ref{eq:recurh2}) from $-\infty$ to $Z$ as
\[
\fancyh_r(Z) \approx \frac{(2r-2)!}{(r-1)!r!}  \cdot \fancyh_1(Z)^{2r-1} 
\]
as justified in Proposition 1(iv).
\item[(ii)]
Working upwards in small steps of $z$, approximate the integral in (\ref{eq:recurh2}) on a grid of points, by the logarithmic trapezium rule. If we write (\ref{eq:recurh2}) as, for short,
\[
I_r(y) = \int_{-\infty}^{y} S_r(z) \, dz
\]
then we have\footnote{The term on the end is the trapezium rule for integrating piecewise exponential functions. Given the typical behaviour of $\fancyh_r(z)$, this is a better idea than linear interpolation.}
\[
I_r(y_{j+1}) \approx I_r(y_j) + (y_{j+1}-y_j) \frac{S_r(y_{j+1})-S_r(y_j)}{\ln S_r(y_{j+1}) - \ln S_r(y_j)} .
\]
\item[(iii)]
Increment $r$ and repeat from (i).
\end{itemize}

\end{algm}

We have said that $\fancyh_r(y)/\fancyh_{r+1}(y)$ gives $\lambda$, in the limit $r\to\infty$, and Algorithm~1 allows the functions to be computed. Empirically, convergence is much faster for $y>0$, and it is desirable to use convergence acceleration techniques when $y<0$. As we will have a fixed boundary in mind, we can write $x_r = \fancyh_r(\barrier)/\fancyh_{r+1}(\barrier)$, a positive sequence that, empirically at least, tends to its limit from above. Write $\delta_r=x_r-x_{r-1}$ for its sequence of differences. 
One of the commonest methods of accelerating convergence is \emph{Aitken's method}, which is that the derived sequence
\begin{equation}
(\aitken_0x)_r := x_r + \frac{\delta_r^2}{\delta_{r-1}-\delta_r}
\label{eq:Aitken}
\end{equation}
often enjoys faster convergence as $(x_r)$, and to the same limit, particularly if the convergence is linear, i.e.\ $\delta_r/\delta_{r-1} \to \mbox{const}$. Indeed, if differences are in geometric progression, then $(\aitken_0x)_r$ converges immediately, i.e.\ for all $r$ one has $(\aitken_0x)_r=\lim_{n\to\infty} x_n$.

But in our case the sequence $(x_r)$ does not converge linearly, and Aitken's method does not work as well. In fact, the sequence appears to converge logarithmically, i.e.\ with error $O(1/r)$. In particular, as $\barrier\to-\infty$ we are led to study, from Prop.~\ref{prop:1}(iv), the ratio of adjacent Catalan numbers
\[
x_r=\frac{\catalan_{r-1}}{\catalan_r} = \frac{1}{4} + \frac{3}{4(2r-1)} ,
\]
and ask what variant of Aitken's method will give immediate convergence when $x_r = \lambda + (\alpha+\beta r)\inv$ for constants $\alpha,\beta$.
The answer is
\begin{equation}
(\aitken_1x)_r := x_r + \frac{\delta_r(\delta_r+\delta_{r-1})}{\delta_{r-1}-\delta_r} .
\label{eq:Aitken2}
\end{equation}
For the proof, note that the estimate of the limit based on $x_{r-2},x_{r-1},x_r$---we are calling this $(\aitken_1 x)_r$---must be the value $\xi$  such that $(x_{r-2}-\xi)\inv,(x_{r-1}-\xi)\inv,(x_r-\xi)\inv$ lie in arithmetic progression, and solve for $\xi$.
To see how this works for the sequence $\catalan_{r-1}/\catalan_r$, we note that the first few terms of the sequence are
\[
1/1, \; 1/2, \; 2/5, \ldots
\]
and so the first term of the accelerated sequence $(\aitken_1 x)$ is
\[
{\textstyle\frac{2}{5}} + \frac{(\frac{2}{5}-\frac{1}{2})(\frac{2}{5}-\frac{1}{2}+\frac{1}{2}-1)}{(\frac{1}{2}-1)-(\frac{2}{5}-\frac{1}{2})} = {\textstyle\frac{1}{4}}
\]
which is the exact limit---as expected in view of the derivation.
Applied to our problem, we find empirically that (\ref{eq:Aitken2}) works very much better than (\ref{eq:Aitken}), mainly because (\ref{eq:Aitken}) undercorrects.
A general caveat should be mentioned: regardless of what convergence accelerator is applied, numerical instability can result if $|\delta_{r-1}-\delta_r|$ becomes too small, and certainly if it becomes comparable with the machine precision.

\notthis{
Regardless of what convergence accelerator is applied, numerical instability can result if $\delta_{r-1}-\delta_r$ becomes too small, and certainly if it becomes comparable with the machine precision.
In the case at hand, $x_r$ is a decreasing sequence of positive numbers (*** we have not shown that they are decreasing! ***). This inspires the regularised formula
\begin{equation}
(\aitken_1x)_r := x_r + \frac{\delta_r(\delta_r+\delta_{r-1})}{\min(\delta_{r-1}-\delta_r, -\varepsilon x_r)}
\label{eq:Aitken2r}
\end{equation}
where we have taken the dimensionless parameter $\varepsilon=0.001$. Note that this choice relies on the $(x_r)$ being decreasing and positive, so it is suitable for the problem at hand, but not for the general problem of accelerating logarithmically-convergent sequences.
} % end

\subsection{Examples}

We demonstrate Algorithm~1 in several different cases\footnote{We used $Z=-10$ and a grid spacing of $\frac{1}{32}$.}. We only show the first term of the accelerated sequence, and this  requires two previous differences, so we need $\fancyh_1/\fancyh_2$, $\fancyh_2/\fancyh_3$, $\fancyh_3/\fancyh_4$ to compute it.

\begin{figure}[!htbp]
\centering
\begin{tabular}{rl}
(a) &
\scalebox{0.8}{\includegraphics{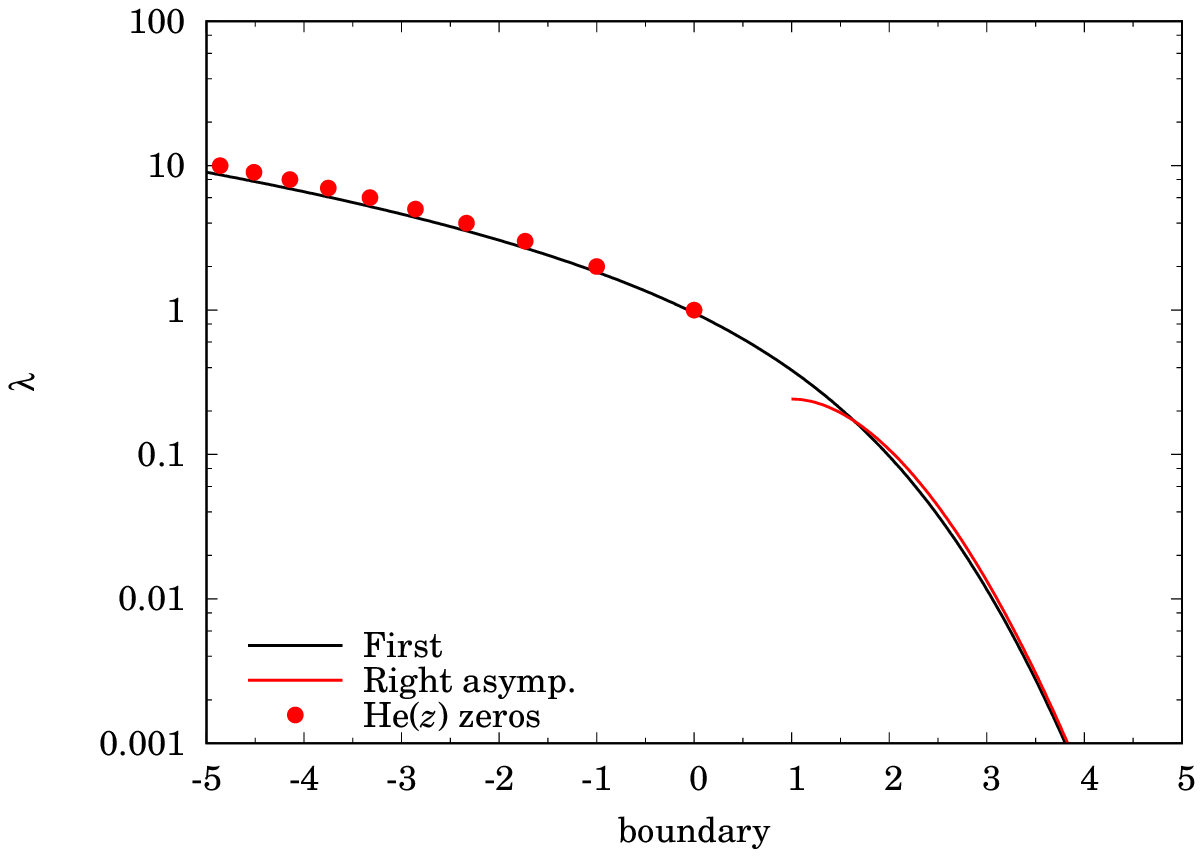}} \\
(b) &
\scalebox{0.8}{\includegraphics{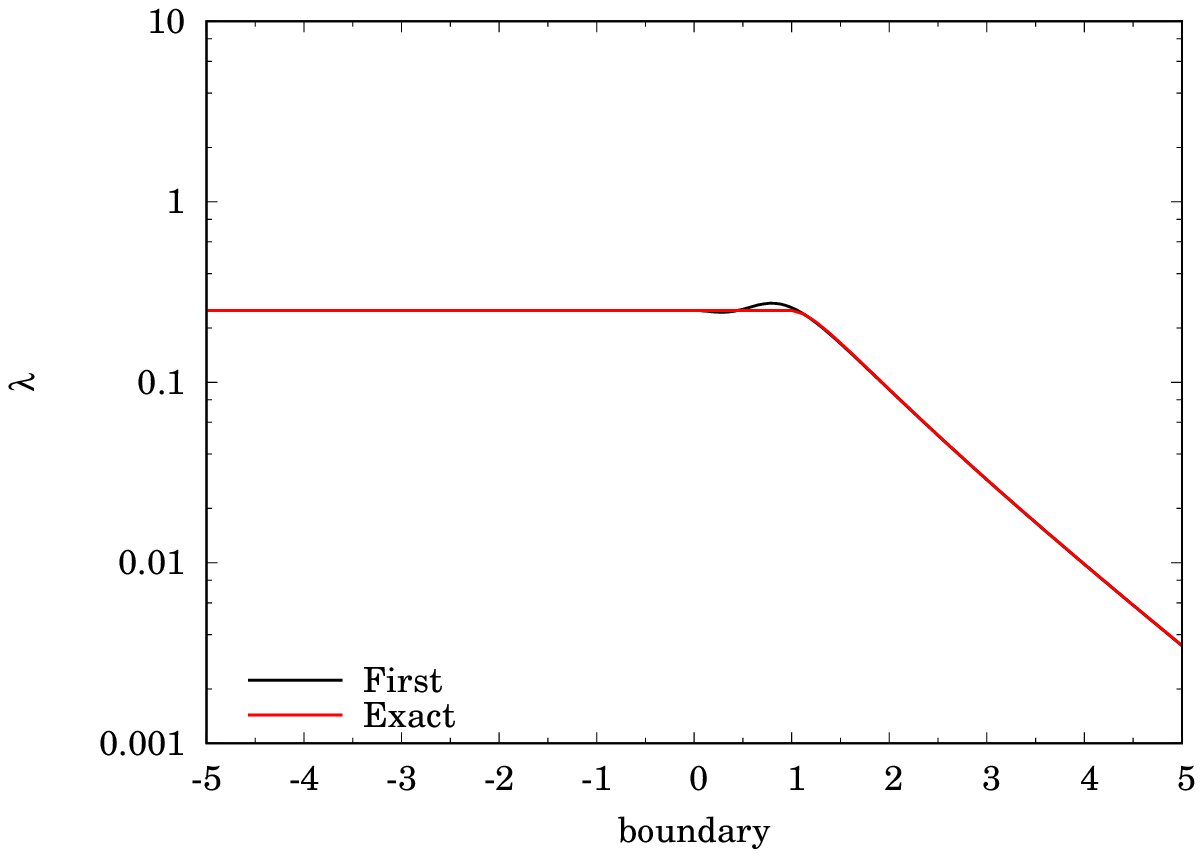}}  \\
(c) &
\scalebox{0.8}{\includegraphics{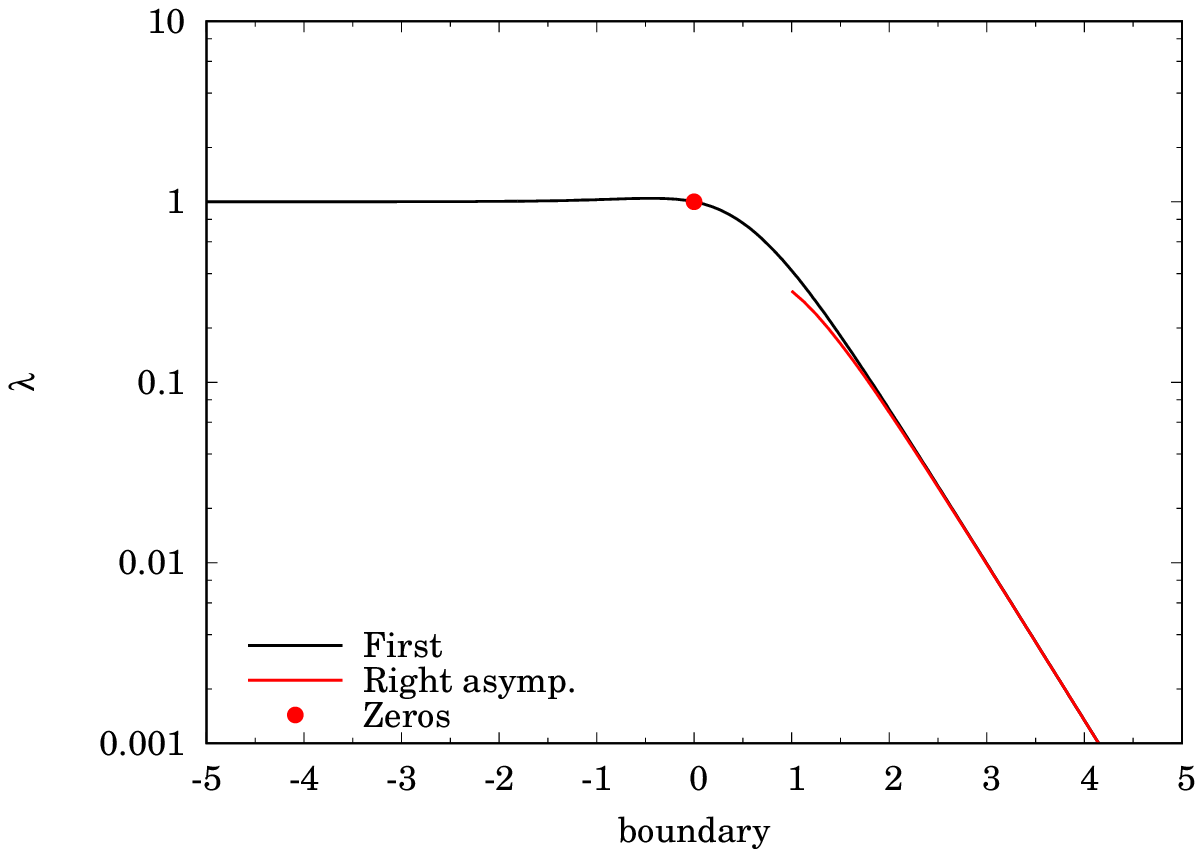}} 
\end{tabular}
\caption{
Performance of Algorithm~1 for (a) OU, (b) Dry-friction, (c) $\ffield(y)=-2\tanh y$. The black line is the estimate of $\lambda$ using the first term of the accelerated sequence.
Where relevant the zeros of the associated orthogonal polynomials are shown, and asymptotes of $\lambda$ vs boundary $\barrier$ are sketched, except in (b) where the exact result is known.
}
\label{fig:1}
\end{figure}

\subsubsection{OU}

Naturally we take this first: $\ffield(y)=-y$, with the results shown in Figure~\ref{fig:1}(a).
The usual approach is to say that we are looking for the zeros of the function $s\mapsto \bigD_s(z)$. See also \cite[Fig.1]{Ricciardi88} and \cite[Fig.1]{Elbert08}, and also Table~\ref{tab:1} in the Appendix which gives some particular values.
In the special case $-s=n\in\N$ this boils down to finding the leftmost zero $\zeta_n$ say of the function $z\mapsto\He_n(z)$, or equivalently the leftmost abscissa of the Gauss-Hermite quadrature formula of order $n$: in other words if the boundary is placed at $\zeta_n$ then $\lambda=n$, and these are plotted in Figure~\ref{fig:1}(a).

The asymptotic behaviour as $\barrier\to -\infty,+\infty$ is, from Theorem~\ref{thm:R3}:
\begin{equation}
\lambda\sim \barrier^2/4 \quad \mbox{ and }  \quad \lambda \sim \barrier \phi(\barrier).
\end{equation}
We can also establish the $y\to+\infty$ result by looking at $\bigD_{-\lambda}(y)$ in this r\'egime (for $\lambda\approx0$). Indeed, by (\ref{eq:pcfrecur}),
\[
\bigD_{-\lambda}(\barrier) = \frac{-\bigD_{-1-\lambda}(\barrier) - \lambda \bigD_{1-\lambda}(\barrier)}{\barrier} ;
\]
now let $\lambda\to0$ and use the expressions for $\bigD_{-1}(\barrier)$ and $\bigD_1(\barrier)$ to get
\[
\bigD_{-\lambda}(\barrier) \approx \frac{\barrier - \lambda \Phi(\barrier)/\phi(\barrier)} {\barrier}
\]
which is zero when $\lambda= \barrier \phi(\barrier)$ (as $\Phi(\barrier)\to1$).
We can derive the same result using (\ref{eq:pcfrecur2}).

At this point we can clearly see why there is something fundamentally wrong with (\ref{eq:wrong}): the decay rate is incorrect, except when $\barrier=0$.

\subsubsection{Arithmetic Brownian motion}

The case $\ffield(y)= \mu$ might seem an odd choice, because it is not mean-reverting, and $\psi(y)$ is formally $e^{\mu y}$ which is not normalisable.
However, provided $\mu>0$, we do have $\ffield\in\mathfrak{S}_-$, and some of the results carry over; also Algorithm~1 does work despite the fact that $\psi(y)$ is not normalisable.
It is easily established that $H(s,y)= (\mu-\sqrt{\mu^2+4s})/2$, regardless of $y$, and Taylor series expansion shows that $\fancyh_r(y)=\catalan_{r-1}\mu^{1-2r}$, for $r\ge1$. This confirms that $\lambda=\mu^2/4$ regardless of the boundary position. 
This case is a useful ansatz for understanding any problem in which $\lim_{y\to-\infty} \ffield(y)=\mu>0$.

\subsubsection{Dry-friction}

Here $\ffield(y)=-\mu \,\sgn y$, with $\mu>0$ (see \cite{Touchette10a}).
When $y\le0$ we have the arithmetic Brownian motion, so then  $\lambda=\mu^2/4$.
For $y\ge0$,
\[
C_+(s,y) = \frac{\sqrt{\mu^2+4s}-\mu}{\sqrt{\mu^2+4s}} e^{(\mu+\sqrt{\mu^2+4s})y/2} + \frac{\mu}{\sqrt{\mu^2+4s}} e^{(\mu-\sqrt{\mu^2+4s})y/2} .
\]
The singularities in the $s$-plane of $H(s,\barrier)$ are a branch-point at $s=-\mu^2/4$ and also simple poles whenever the following condition is satisfied:
\begin{equation}
\sqrt{1+4s/\mu^2} = 1 - e^{-\sqrt{\mu^2+4s} \, \barrier} .
\label{eq:lambda_df}
\end{equation}
When $\barrier<1/\mu$ this has no solutions for $s>{-\mu^2/4}$, so $\lambda$ is still equal to $\mu^2/4$.
When $\barrier>1/\mu$ it has two in the interval $({-\mu^2/4},0)$ and the important (rightmost) root is at $s=-\lambda$ obeying
\[
\lambda \sim \frac{\mu^2 e^{-\mu y_+}}{2}, \qquad \barrier+\to+\infty.
\]
As $\psi(y)=\mu e^{-\mu |y|}/2$, this accords with Theorem~\ref{thm:R3}. 
In running examples we can take $\mu=1$ without loss of generality and Figure~\ref{fig:1}(b) shows the results, using (\ref{eq:lambda_df}) as a check.

\subsubsection{tanh case}

The case $A(y)=-\alpha \tanh \gamma y$ is an interesting generalisation, interpolating between the OU and dry-friction cases.

The function $C_+(s,y)$ obeys
\[
\dderiv{C}{y} - \alpha \tanh \gamma y \deriv{C}{y} - sC = 0
\]
and substituting $w=\sinh \gamma y$ turns the equation for $C$ into the hypergeometric equation
\[
(1+ w^2) C''(w) + (1-\alpha/\gamma) wC'(w) - (s/\gamma^2) C=0.
\]
Polynomial solutions are admitted for certain values of $s=-\lambda_n$, as detailed below for the first few:
\begin{center}
\begin{tabular}{rrr}
$n$ & $P_n(w)$  & $\lambda_n$ \\ \hline
1 & $w$ & $\gamma(\alpha-\gamma)$ \\
2 & $w^2 + \gamma/(2\gamma-\alpha)$ & $2\gamma(\alpha-2\gamma)$ \\
3 & $w^3 + 3\gamma w/(4\gamma-\alpha)$ & $3\gamma(\alpha-3\gamma)$
\end{tabular}
\end{center}
% Not sure we need Raposo ref as well
These are the Romanovski polynomials, the orthogonal polynomials of the Student-t distribution \cite{Quesne13,Romanovski29}, but the class is defective in the sense that there are only finitely many of them. They are in the Pearson-Wong family \cite{Wong64} but are usually omitted in discussions of the subject, probably because of their irregular behaviour.
The apparent pattern for the $(\lambda_n)$ is confirmed by the usual three-term recurrence relation of orthogonal polynomials (e.g.~\cite{Szego39}), which yields
\[
\lambda_{n+1} = \lambda_{n} + \gamma(\alpha-\gamma) - 2\gamma^2 n . 
\]
If $\alpha\to\infty$, $\gamma\to0$ with $\alpha\gamma=1$ then we are back with OU and the $(P_n)$ become Hermites. 
If $\alpha\to\infty$ with $\gamma$ fixed then we have the dry-friction case and the zeros form a continuum.

The defectiveness of the set of Romanovski polynomials has important practical consequences. Take for example $\ffield(y)=-2\tanh y$, so that $\psi(y) = \half \sech^2 y$. For $\alpha/\gamma=2$, the set of polynomials terminates even before $n=2$, as formally $P_2(w)=w^2+\infty$; put differently, $P_2$ is a polynomial with two real zeros only when $\alpha/\gamma>2$. In identifying points on the graph of $\lambda$ vs $y$ as zeros of orthogonal polynomials, we can only plot the zero of $P_1(w)$ before we get stuck. Other aspects of what we have derived can, however, be easily plotted. Indeed, as $y\to-\infty$,  by reference to the arithmetic Brownian motion case, we have $\lambda\to1$, and as $\barrier\to+\infty$ we have $\lambda\sim\sinh \barrier/\cosh^3 \barrier$ (see Figure~\ref{fig:1}(c)).

%%%%%%%%%%%%%%%%%%%%%%%%%%%%%%%%%%%%%%%%%%%%%%%%%%%%%%%%%%%%%%
%%%%%%%%%%%%%%%%%%%%%%%%%%%%%%%%%%%%%%%%%%%%%%%%%%%%%%%%%%%%%%
%%%%%%%%%%%%%%%%%%%%%%%%%%%%%%%%%%%%%%%%%%%%%%%%%%%%%%%%%%%%%%
%%%%%%%%%%%%%%%%%%%%%%%%%%%%%%%%%%%%%%%%%%%%%%%%%%%%%%%%%%%%%%

\section{Short-time behaviour, and global asymptotics}
\label{sec:short}

\subsection{General theory and limiting behaviours}

Another branch of the theory---following on from \cite{Martin15b}---is to study the logarithmic derivative of the density.
Our analysis is guided, in part, by the first passage problem for the regular Brownian motion, i.e.\ with no mean reversion (drift term just $\mu\,dt$). For that problem, using $(t,x)$ coordinates,
\[
F(t,x) = \Phi\!\left( \frac{\mu t +x-\barrierx}{2\sigma\!\sqrt{t}} \right) + e^{2\mu (\barrierx-x)/\sigma^2} \Phi\!\left( \frac{-\mu t +x- \barrierx}{2\sigma\!\sqrt{t}} \right)
\]
or
\[
f(t,x) = \pderiv{F}{t}(t,x) =  \frac{\barrierx-x}{\sqrt{2\pi \sigma^2 t^3}} \exp \left( \frac{-(\barrierx-x-\mu t)^2}{2\sigma^2 t} \right) ,
\]
the well-known inverse Gaussian distribution \cite{Seshadri93}.
Now consider the logarithmic derivative w.r.t.\ $x$:
\[
-\pderiv{}{x} \ln f(t,x)
=
\frac{x+\mu t-\barrierx}{\sigma^2 t} + \frac{1}{\barrierx-x} 
\]
which is a particularly simple (rational) function.

In the OU case, (\ref{eq:wrong}) is correct when $\barrier=0$, i.e.\ the boundary is at the equilibrium point, and again the logarithmic derivative is a simple function:
\[
f (\tau,y) = \frac{2y \sqrt{q}}{\sqrt{2\pi(1-q)^3}} \exp \left( \frac{-qy^2}{2(1-q)} \right) ;
\qquad -\pderiv{}{y}\ln f = \frac{qy}{1-q} - \frac{1}{y}.
\]

All this points to the logarithmic derivative being a useful construction for dissecting the problem. We formalise the idea next.
It is convenient to define ($'$ denoting $\partial/\partial y$)
\begin{equation}
\riccati [h] \equiv h' + \ffield h - h^2.
\end{equation}
\begin{prop}
With $h(\tau,y)=-(\partial/\partial y) \ln f(\tau,y)$ we have:
\begin{itemize}

\item[(i)]
\begin{equation}
\pderiv{h}{\tau} = \pderiv{}{y} \riccati[h].
% \left\{\ffield(y) h +  \pderiv{h}{y} - h^2  \right\}.
\label{eq:pde_hy}
\end{equation}

\item[(ii)]
Near $\barrier$ the function $h$ looks like
\begin{equation}
h(\tau,y) = \frac{1}{\barrier-y} + \frac{A(\barrier)}{2} + o(1)
\label{eq:hbarr}
\end{equation}
for all time. 

\item[(iii)]
For short time ($\tau\ll 1$) we have
%, with $q=e^{-2\tau}$,
\begin{equation}
h(\tau,y) = \frac{y-\barrier}{2\tau} + \frac{\ffield(y)}{2} + \frac{1}{\barrier-y} +  o(1)
\label{eq:ansatz1}
\end{equation}

\item[(iv)]
The steady-state solution for $h$, i.e.\ $\hinfty(y)=h(\infty,y)$, obeys
\begin{equation}
\riccati[\hinfty] = \lambda.
\label{eq:hinfty}
\end{equation}

\item[(v)]
Splitting off the part of $\hinfty$ that is singular at the boundary, thereby defining 
\begin{equation}
\hinfty(y) = \frac{1}{\barrier-y} +  \hinftysharp(y),
\label{eq:hsharpbarr}
\end{equation}
we have
\begin{equation}
\hinftysharp(\barrier) = \ffield(\barrier)/2 ;
\label{eq:hsharpbarr0}
\end{equation}
if $A$ is differentiable at $\barrier$ then
\begin{equation}
\hinftysharp'(\barrier) = \displaystyle \frac{1}{3} \left( \lambda - \frac{\ffield(\barrier)^2}{4} + \ffield'(\barrier) \right) .
\label{eq:hsharpbarr1}
\end{equation}
%and if $\ffield(y)=\sum_{r=0}^\infty a_r(y-\barrier)^r$, $\hinftysharp(y) = \sum_{r=0}^\infty h_r^+(y-\barrier)^r$, then
%\begin{equation}
%h_{r+1}^+ = \frac{1}{r+3} \left( a_{r+1} + \sum_{k=0}^r (h_k^+ - a_k) h_{r-k}^+  \right) , \qquad r\ge1.
%\label{eq:hsharpbarr2}
%\end{equation}

\end{itemize}
\end{prop}

\noindent Proof.
(i) Clear, and (ii) clear by dominant balance.
(iii) Again by dominant balance, and corresponding to an approximation by which the process is viewed as an arithmetic Brownian motion over a short time period.
(iv) Clear by taking (\ref{eq:pde_Fy}) (which is obeyed by $f$), dividing through by $f$ and letting $\tau\to\infty$, with $(1/f)\partial f/ \partial \tau \to -\lambda$. Part (v) is also immediate.
$\Box$

\gap

From this we can see something else that is wrong with (\ref{eq:wrong}): it implies that
\[
h(\tau,y) \stackrel{??}{=} \frac{qy-\!\sqrt{q}\,\barrier}{1-q} + \frac{1}{\barrier-y} .
\]
This has incorrect asymptotic behaviour: as $\tau\to\infty$ only the $(\barrier-y)\inv$ term remains, but that does not satisfy the Riccati equation (\ref{eq:hinfty}), except when $\barrier=0$.  It is also incorrect at the boundary in the sense that although the $O(\barrier-y)\inv$ term is correct, the $O(1)$ term is not: it is equal to $\frac{q-\sqrt{q}}{1-q}\barrier$, but should be $-\half\barrier$, which agrees only as $\tau\to0$. Yet it does have, informally, `some of the right terms'.

It is possible to refine (\ref{eq:ansatz1}) by incorporating extra terms in the expansion, again for small $\tau$. This, however, introduces unwelcome complications and is explained in the Appendix. In a nutshell the conclusion is that $h$ does \emph{not} admit a convergent expansion of the form
\[
\frac{y-\barrier}{2\tau} + \frac{\ffield(y)}{2} + \frac{1}{\barrier-y} 
+ \sum_{r=1}^\infty \tau^r b_r(y).
\]
In other words, taking the logarithmic derivative of $f$ does not in general remove the essential singularity of $f$ at $\tau=0$, though as we know it does in some cases, notably the Brownian motion with drift and the OU model with the boundary at equilibrium. We therefore shift our attention away from short-time development, and concentrate on making (\ref{eq:ansatz1}) work in the long-time limit as well.

\subsection{Longer-time development of $h$}

Write $q=e^{-2\theta \tau}$, where $\theta>0$ is an arbitrary constant, and consider the following ansatz:
\begin{equation}
h(\tau,y) = \frac{\theta\!\sqrt{q}\,(y-\barrier)}{1-q} + \frac{\sqrt{q}\,\ffield(y)}{1+\!\sqrt{q}}  + \frac{1}{\barrier-y} + \frac{1-\!\sqrt{q}}{1+\!\sqrt{q}} \, \hinftysharp(y) + R(\tau,y).
\label{eq:h2}
\end{equation}
It is easily seen that:
\begin{itemize}
\item
It has the desired behaviour at the boundary, as given in (\ref{eq:hbarr});
\item
Its Laurent series around $\tau=0$ agrees with that of (\ref{eq:ansatz1}), regardless of $\theta$;
\item
It has the desired long-time behaviour, as is immediate from letting $q\to0$ and recalling (\ref{eq:hsharpbarr});
\item
In the OU case with $\ffield(y)=-\theta y$ and $\barrier=0$ it is exact;
\item
In the case of the arithmetic Brownian motion it is also exact.
\end{itemize}

There are other connections with known results.
First, in the OU model $\ffield(y)=-\theta y$, consider for some fixed $\xi$ the function
\[
\psi_\textrm{b}(\tau,y) = \lim_{\varepsilon\to0} \frac{1}{2\varepsilon} \pr\big[|Y_\tau-\xi|<\varepsilon \,\big|\, Y_0=y\big]
\]
i.e.\ the p.d.f.\ of the unconstrained process, which is a solution to the \emph{backward} equation (because the spatial variable $y$ is the initial condition). If $h=-(\partial/\partial y)\ln \psi_\textrm{b}$ then
$h(\tau,y)=\frac{\theta(qy-\sqrt{q}\xi)}{1-q}$. So the first two terms of (\ref{eq:h2}), taken together, provide a solution satisfying (\ref{eq:pde_hy}), but it is not one that we can directly use as it ignores the boundary.
The third and fourth terms can, respectively, be understood as introducing the effect of the absorption and ensuring that the right behaviour is observed in the long-time limit.

A second connection is with the work in \cite{Martin18b} which deals with the Fokker--Planck equation for general mean-reverting processes with no absorbing boundary. Define
\[
\psi_\textrm{f}(\tau,y) = \lim_{\varepsilon\to0} \frac{1}{2\varepsilon} \pr\big[|Y_\tau-y|<\varepsilon \,\big|\, Y_0=\xi\big]
\]
which satisfies the Fokker--Planck equation. Then\footnote{Noting that $\psi(y)$ as previously defined is simply $\psi_\textrm{f}(\infty,y)$.}  $\psi_\textrm{f}(\tau,y)/\psi(y)$ is a solution of the backward equation and hence its logarithmic derivative satisfies (\ref{eq:pde_hy}). In \cite{Martin18b} the first two terms of (\ref{eq:h2}) are used to provide an approximate solution to the Fokker--Planck equation for general mean-reverting models. The idea is to expand the Fokker--Planck problem `around' the OU model to which it is closest, and the solution is exact for $\ffield(y)=-\theta y$. (We reiterate that in \cite{Martin18b} there is no absorbing boundary.)

In summary, from the way that (\ref{eq:h2}) is constructed, we have:
\begin{itemize}
\item
$R(\tau,y)\to0$ as $\tau\to0,\infty$;
\item
$R(\tau,\barrier)=0$;
\item
$R$ vanishes in (i) the OU case with the boundary at equilibrium and (ii) the arithmetic Brownian motion (OU with no mean reversion).
\end{itemize}

We have therefore made an important step in constructing an approximation that is valid over short and long time scales. The connection with the first section of the paper is that $\lambda$ has appeared in (\ref{eq:hsharpbarr1}). In principle it is possible to identify $\lambda$ by solving the eigenvalue problem (\ref{eq:hinfty}), which would then render the work in \S\ref{sec:lambda} unnecessary. However the solution of (\ref{eq:hinfty}) is not straightforward. Furthermore, it transpires that it is not necessary to know $\hinftysharp$ to make further progress, if we are just interested in the behaviour of $f$ over time (as opposed to as a function of the starting-point for fixed time).

\subsection{Choice of $\theta$}

We said above that $\theta$ is arbitrary, and its effect on the Laurent series of $h$ about $\tau=0$ is confined to the $o(1)$ term in (\ref{eq:ansatz1}), so it controls the \emph{intermediate}-time behaviour.
Given the connection with the OU process as described above, the role of $\theta$ is to map the given force-field $\ffield$ on to its `closest' OU model in some sense.

In \cite{Martin15b} we suggested using $\hat\theta$ defined as the average rate of mean reversion, defined as the average of $-\ffield'$ over the invariant density $\psi$:
\begin{equation}
\hat{\theta} = \langle -\ffield' \rangle_\infty = \langle \ffield^2 \rangle_\infty
\label{eq:theta}
\end{equation}
and this identity shows that $\hat{\theta}$ is necessarily positive.
As pointed out in \cite{Martin18b} this choice corresponds to the Fisher information (see e.g.~\cite[\S2.5]{Lehmann98}) for the problem of estimating the mean by maximum likelihood. More precisely, consider for some p.d.f.~$\psi$ the family of distributions $\psi(y-\mu)$ indexed by the parameter $\mu\in\R$. Writing
\[
f(y \cdl \mu) = \psi(y-\mu)
\]
we seek to maximise $\log f(y \cdl \mu)$ w.r.t.\ $\mu$. The Fisher information is the expectation of the square of the $\mu$-derivative of the log-likelihood, and hence is 
\[
\int_{-\infty}^\infty \left(\pderiv{}{\mu} \log f(y\cdl\mu)\right)^2 f(y\cdl\mu) \, dy 
=
\int_{-\infty}^\infty \left(\frac{\psi'(y-\mu)}{\psi(y-\mu)} \right)^2 \psi(y-\mu) \, dy
=
\int_{-\infty}^\infty \left(\frac{\psi'(y)}{\psi(y)} \right)^2 \psi(y) \, dy
= \hat{\theta}.
\]
In broad terms, the higher the Fisher information, the more certain we are about the estimation of the parameter in question. The connection with mean reversion is that the higher the average speed of mean reversion, the more certain we are about our estimate of the mean from a given dataset, and vice versa. Using the Fisher information as an estimator of reversion speed is therefore natural.
%As also pointed out in \cite{Martin18b}, the definition extends to multiple dimensions: then, the Fisher information is a positive definite symmetric matrix rather than just a positive number. 

For example the tanh case has (with $\Beta$ denoting the Beta function)
\[
\ffield(y)=-\frac{\alpha}{\gamma}\tanh\gamma y
, \qquad
\psi(y) = \frac{\gamma(\cosh\gamma y)^{-\alpha/\gamma^2}}{\Beta\big(\frac{\alpha}{2\gamma^2},\half\big)}, \qquad
\langle-\ffield'\rangle_{\infty}=\frac{\alpha^2}{\alpha+\gamma^2}.
\]

\subsection{Long- and short-time behaviour combined}

The present state of affairs is that we know the asymptotic rate of decay in the long-time limit ($\lambda$). Also we know a fair amount about the logarithmic derivative of $f$, i.e.\ $h(\tau,y)$; but this only allows us to reconstruct $f(\tau,y)$ up to a multiplicative time-dependent factor $\gothicn(\tau)$ say, which we must now obtain. Symbolically
\begin{equation}
f(\tau,y) = \gothicn(\tau) \exp \left( -\int_{y_*}^y h(\tau,z) \, dz \right);
\label{eq:decomp}
\end{equation}
the lower limit $y_*$ of the integral is arbitrary.

We calculate the exponential-integral in (\ref{eq:decomp}) first, using (\ref{eq:h2}) and defining
\[
\kcoefstuff = \int_y^{\barrier} \hinftysharp(z) \, dz
\]
to give
\[
\frac{\barrier-y}{\barrier-y_*} 
\exp \left( \frac{\theta\sqrt{q}}{1-q} \frac{(y_*-y_+)^2}{2} \right)
\exp \left( - \frac{\theta\sqrt{q}}{1-q} \frac{(y-y_+)^2}{2} \right)
\left( \frac{\psi(y_*)}{\psi(y)} \right)^{ \textstyle \frac{\sqrt{q}}{1+\sqrt{q}} }
e^ {\kcoefstuff \textstyle \frac{1-\sqrt{q}}{1+\sqrt{q}} } .
\]
The prefactor $(\barrier-y_*)\inv$ can be absorbed into the $\gothicn(\tau)$ term, which means that in effect we can discard it. This permits us to let $y_*\to\barrier$, giving:
\[
(\barrier-y)
\exp \left( \frac{- \theta\sqrt{q}(y-\barrier)^2}{2(1-q)} \right)
\left( \frac{\psi(y_+)}{\psi(y)} \right)^{ \textstyle \frac{\sqrt{q}}{1+\sqrt{q}} }
e^ { \kcoefstuff \textstyle \frac{1-\sqrt{q}}{1+\sqrt{q}} } .
\]

We now turn to the prefactor $\gothicn(\tau)$. Inserting (\ref{eq:decomp}) into (\ref{eq:pde_hy}) gives a first-order linear differential equation for $\gothicn$:
\[
- \frac{1}{\gothicn} \deriv{\gothicn}{\tau} = 
\riccati[h] - \int_{y_*}^y  \dot{h}(\tau,z) \, dz ,
\]
with $\dot{h}=\partial h/\partial \tau$.
Notice that the RHS seems to depend on $y$, but does not do so, because $h_Y$ obeys (\ref{eq:pde_hy}). Thus any $y$ can be chosen, and setting it equal to $y_*$ causes the second term to vanish. Then we let $y_*\to\barrier$ to obtain
\begin{equation}
\gothicn(\tau) = \exp \left( - \int_\cdot^\tau  \riccati[h](\tau,\barrier) \, d\tau \right) \times \mbox{const}
\label{eq:prefac}
\end{equation}
where the lower integration limit is arbitrary and only influences the multiplicative constant.
Using (\ref{eq:hsharpbarr0},\ref{eq:hsharpbarr1},\ref{eq:h2}):
\[
\riccati[h](\tau,\barrier) = 
\frac{3\theta q}{1-q} + \lambda + \frac{\theta\ecoef\sqrt{q}}{1+\!\sqrt{q}} + o(R) 
\]
where the constant $\ecoef$ is defined by
\begin{equation}
\theta\ecoef = 3\theta - 2\lambda + \ffield'(\barrier) + \frac{\ffield(\barrier)^2}{2}
\label{eq:ecoef}
\end{equation}
and the symbol $o(R)$ denotes a function that vanishes if  $R$ is identically zero.
Doing the $\tau$-integral, recalling $d\tau=-dq/2\theta q$, gives
\[
M(\tau) = C(1-q)^{-3/2} q^{\lambda/2\theta} \left(\frac{1+\!\sqrt{q}}{2}\right)^{\ecoef}
\]
($C$ denotes a positive constant) or equivalently
\[
M(\tau) = \frac{Ce^{-\lambda \tau}}{(1-e^{-2\tau})^{3/2}} 
\left(\frac{1+e^{-\tau}}{2}\right)^{\ecoef}
\]
% \ecoef = 2(1-\lambda)+\barrier^2/2 for OU
and two important ingredients can be seen: the scaling law for short time is  $\propto\tau^{-3/2}$, seen from the Brownian motion approximation, and the asymptotic decay rate is $\lambda$, as it must be.
 The former is clearly visible in (\ref{eq:wrong}) but the latter is not.

We are now ready to combine it with the previous working to give
\begin{eqnarray}
f(\tau,y) &\approx&
\frac{C(\barrier-y)e^{-\lambda \tau}}{\sqrt{(1-q)^3}}  
\exp \left( \frac{- \theta\sqrt{q}(y-\barrier)^2}{2(1-q)} \right)
\label{eq:almostfinal}
 \\
 && \times 
\left( \frac{\psi(y_+)}{\psi(y)} \right)^{ \textstyle \frac{\sqrt{q}}{1+\sqrt{q}} }
\left(\frac{1+\sqrt{q}}{2}\right)^{\ecoef} 
e^{\kcoefstuff \textstyle \frac{1-\sqrt{q}}{1+\sqrt{q}}} .
\nonumber
\end{eqnarray}
To determine the overall scaling factor $C$ we consider what happens when the process starts near the boundary by setting $y=\barrier-\varepsilon$ and allowing $\varepsilon\to0$. The density must integrate to unity and, making the substitution\footnote{Not $u=\varepsilon^2 \theta \!\sqrt{q}/(1-q)$. The term in the exponential can can be manipulated as $\sqrt{q}/(1-q) = q/(1-q) + \half + o(1)$ as $q\to1$.} $u=\varepsilon^2 \theta q/(1-q)$, we obtain
\[
\int_0^\infty f(\tau,y) \, d\tau = \frac{C}{2\theta^{3/2}} \int_0^\infty u^{1/2} e^{-u/2} (\ldots) \, du
\]
where the expression $(\ldots)$ tends to unity as $\varepsilon\to0$; notice that in this limit all three terms in the second line of (\ref{eq:almostfinal}) disappear, essentially because we are only interested in $y \approx \barrier$ and $\theta\tau\ll1$.
Therefore $C=\sqrt{2\theta^3/\pi}$ and we arrive at 
\begin{equation}
f(\tau,y) \approx
\frac{(\barrier-y)e^{-\lambda \tau}}{\sqrt{\pi(1-q)^3/2\theta^3}}  
\exp \left( \frac{- \theta\sqrt{q}(y-\barrier)^2}{2(1-q)} \right)
\left( \frac{\psi(y_+)}{\psi(y)} \right)^{ \textstyle \frac{\sqrt{q}}{1+\sqrt{q}} }  
\left(\frac{1+\sqrt{q}}{2}\right)^{\ecoef} e^{\kcoefstuff \textstyle \frac{1-\sqrt{q}}{1+\sqrt{q}}} ,
\label{eq:final}
\end{equation}
with $\ecoef$ given by (\ref{eq:ecoef}).
Although we still do not know $\kcoefstuff$, its value can be ascertained by requiring $\int_0^\infty f(\tau,y) \, d\tau=1$, which can be done by Gaussian quadratures and a numerical bisection search \cite{NRC}, and with this final step we are done\footnote{It is perhaps odd at first sight that we are using this principle twice, to obtain two different pieces of information: $C$ and $\kcoefstuff$. The point is that $C$, above, is governed by the short-term behaviour and  the limit $y\to\barrier$ screens out any terms that pertain to long-term behaviour, as the process will hit the boundary almost immediately in that limit.}.

\subsection{Examples}

\subsubsection{OU process}

It is easily seen that in the special case of the OU process with the boundary at equilibrium, we have $\psi(y)=(2\pi)^{-1/2} e^{-y^2/2}$, $\lambda=1$ and $\ecoef=0$, and it is easily seen that $\kcoefstuff=0$ too\footnote{By integrating using essentially the same substitution as before, $u=q/(1-q)$.}, so that (\ref{eq:wrong}) is recovered. More subtly it identifies why and how (\ref{eq:wrong}) is incorrect whenever $\barrier\ne0$.

\begin{figure}[!htbp]
\hspace{-10mm}
\begin{tabular}{llll}
(i) & \scalebox{0.6}{\includegraphics{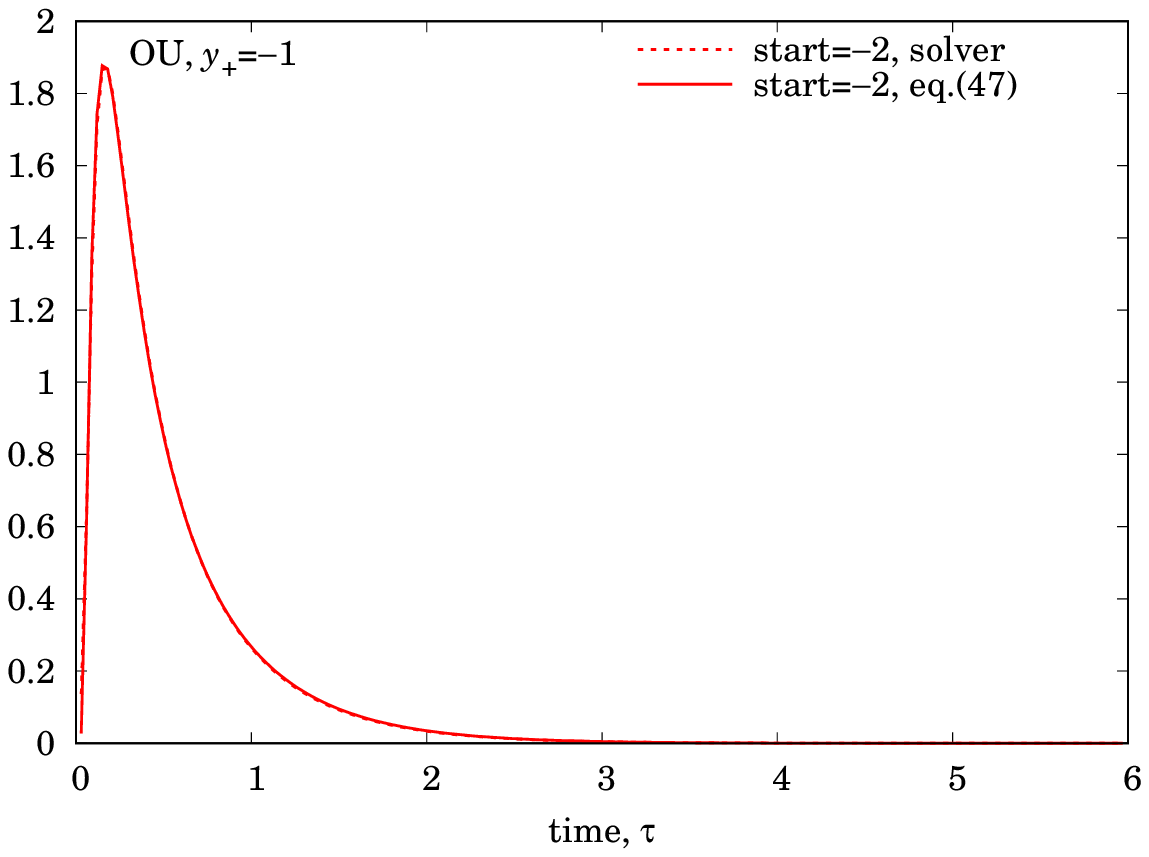}} &
(ii) & \scalebox{0.6}{\includegraphics{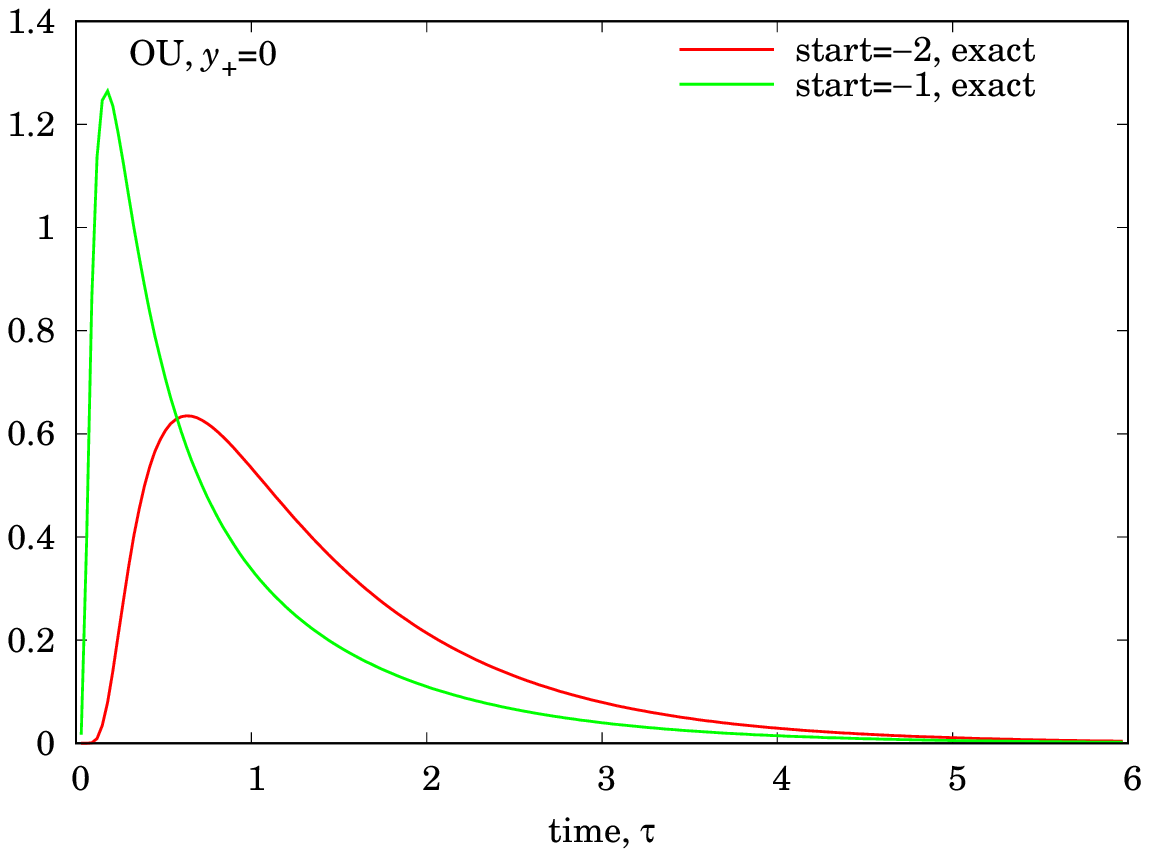}} \\
(iii) & \scalebox{0.6}{\includegraphics{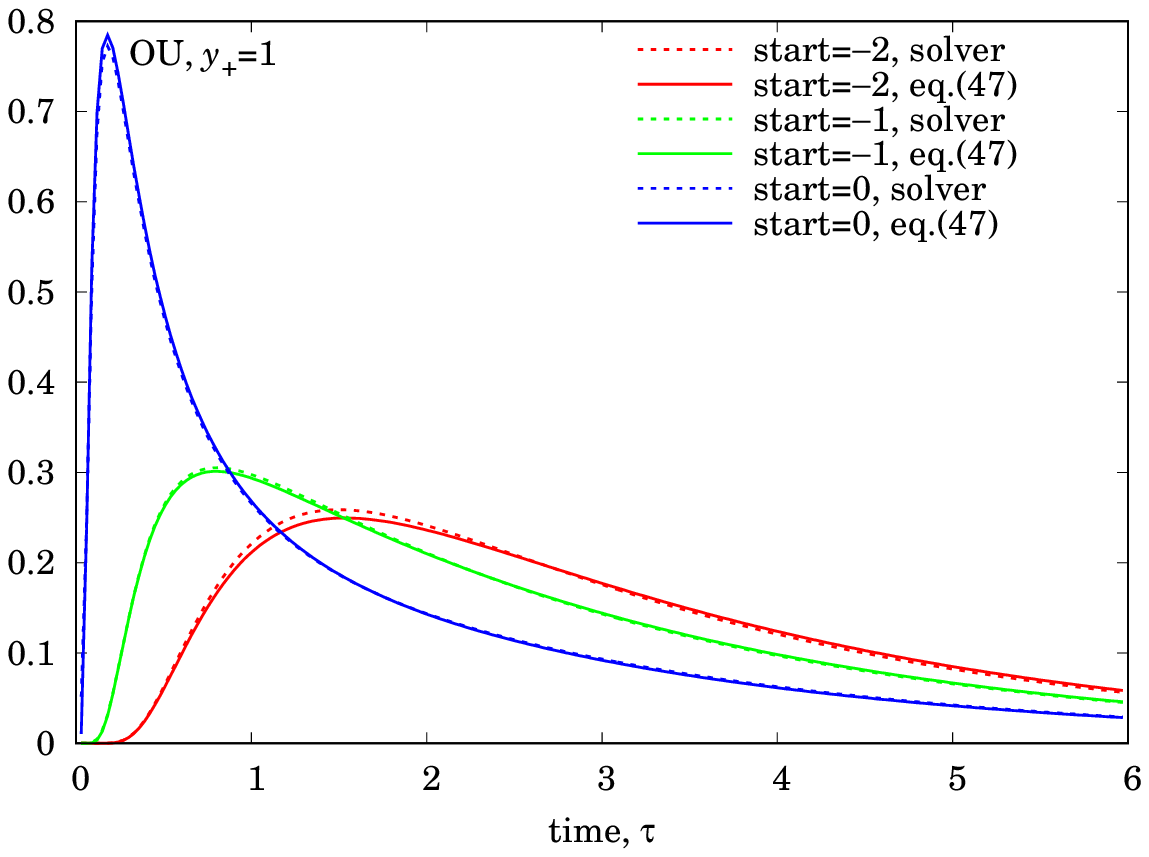}} &
(iv) & \scalebox{0.6}{\includegraphics{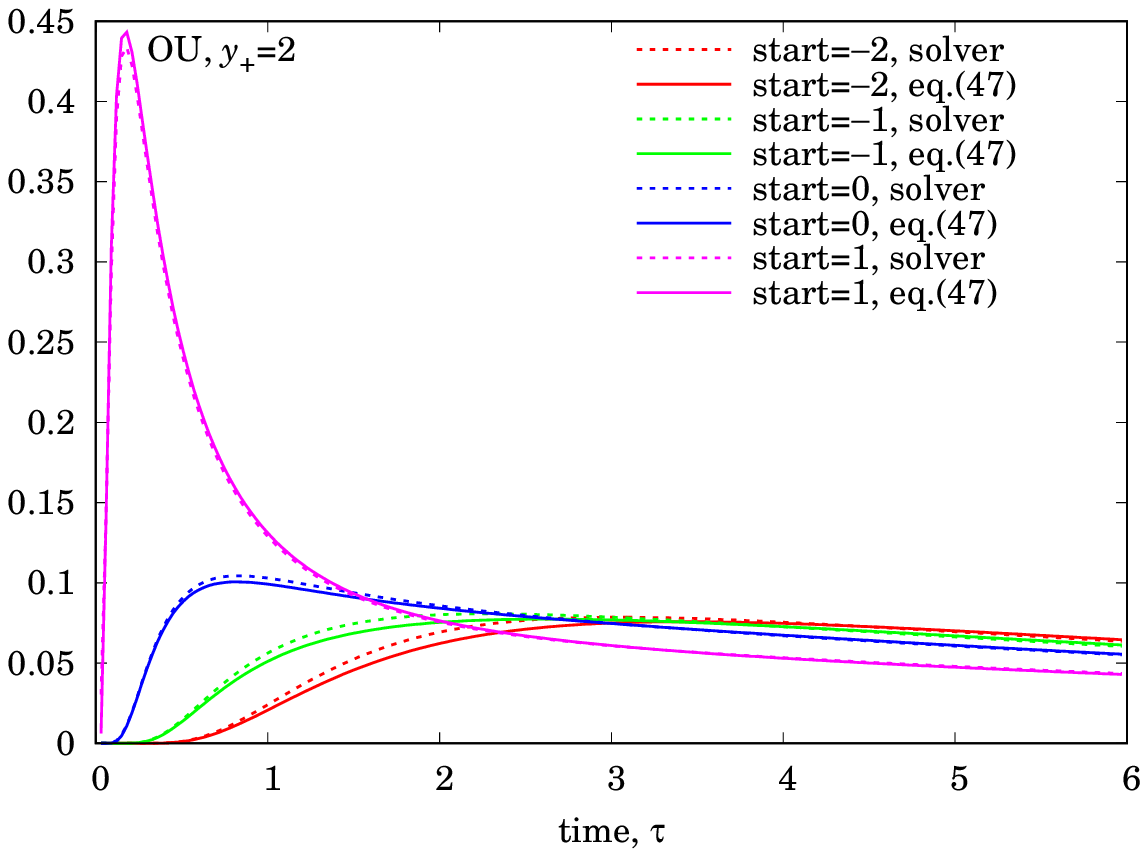}} \\
\end{tabular}
\caption{
First-passage time density for $\ffield(y)=-y$ (OU): Eq.(\ref{eq:final}) compared with numerical PDE solver, except for $\barrier=0$, when (\ref{eq:final}) is exact.  
Boundaries and starting-points as indicated on each plot.
}
\label{fig:ou}
\end{figure}

To investigate the accuracy of (\ref{eq:final}) we use a numerical PDE solver.
Figure~\ref{fig:ou} shows the results for OU with the boundary at different positions, using various starting-points for each. The agreement is very good.

\begin{figure}[!htbp]
\hspace{-10mm}
\begin{tabular}{llll}
(i) & \scalebox{0.6}{\includegraphics{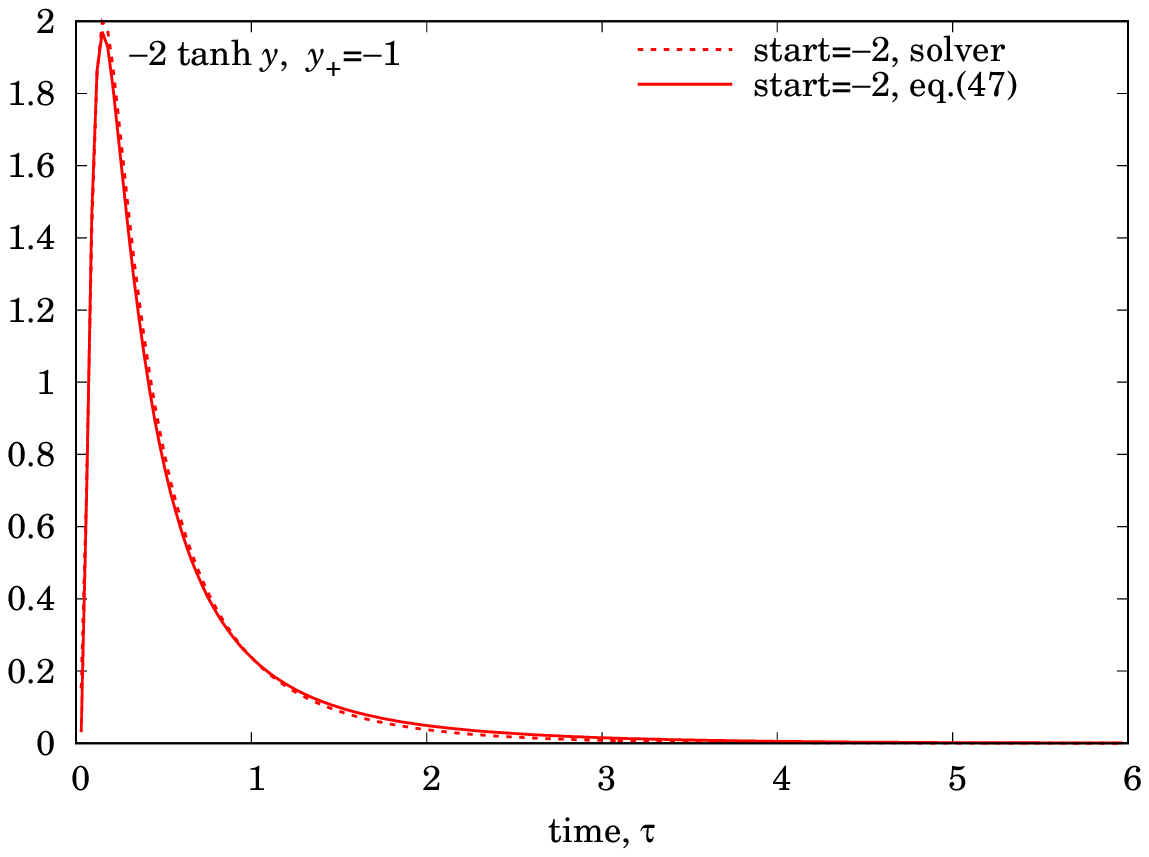}} &
(ii) & \scalebox{0.6}{\includegraphics{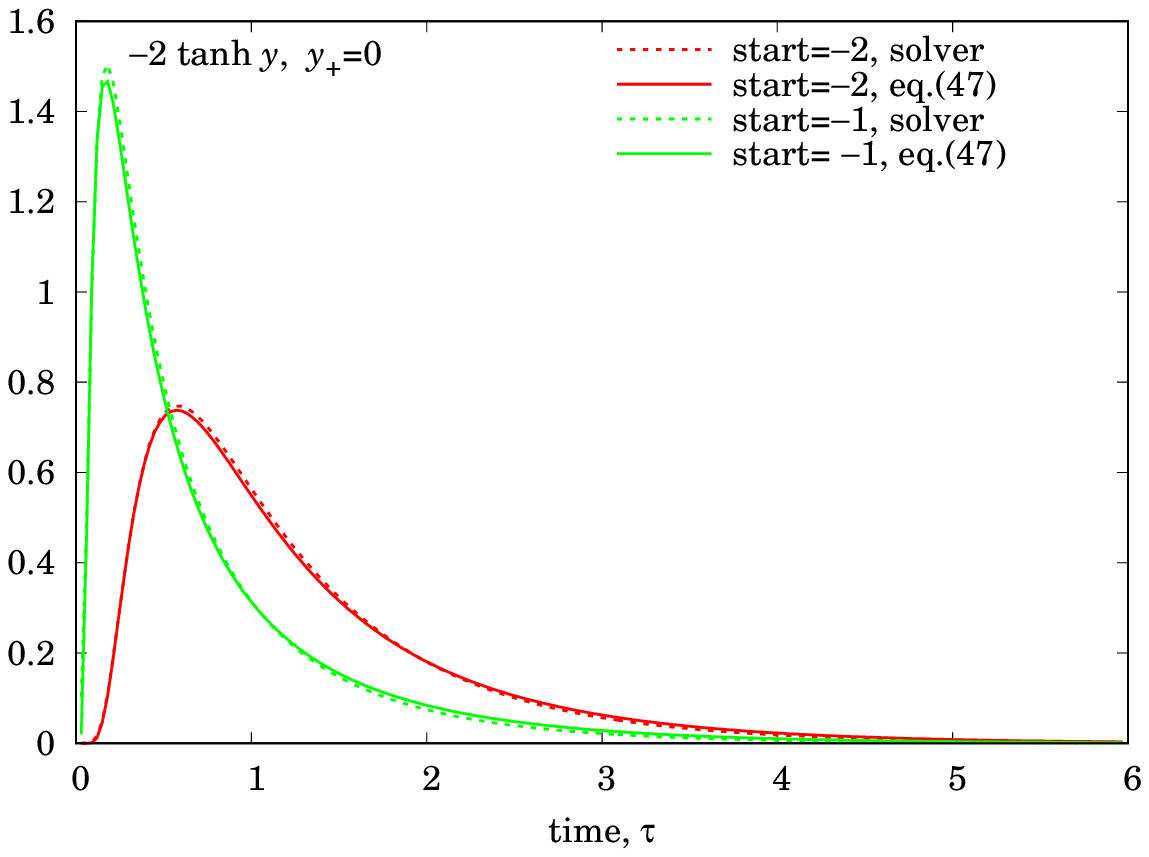}} \\
(iii) & \scalebox{0.6}{\includegraphics{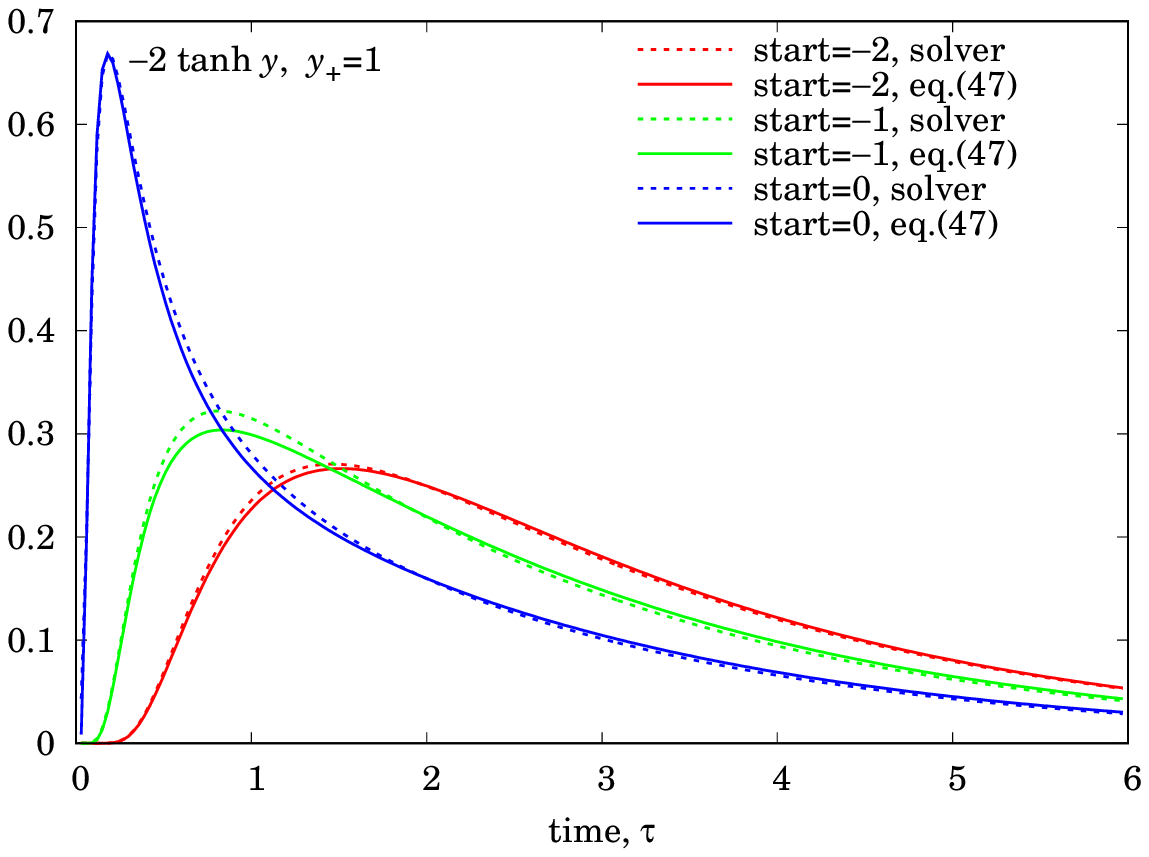}} &
(iv) & \scalebox{0.6}{\includegraphics{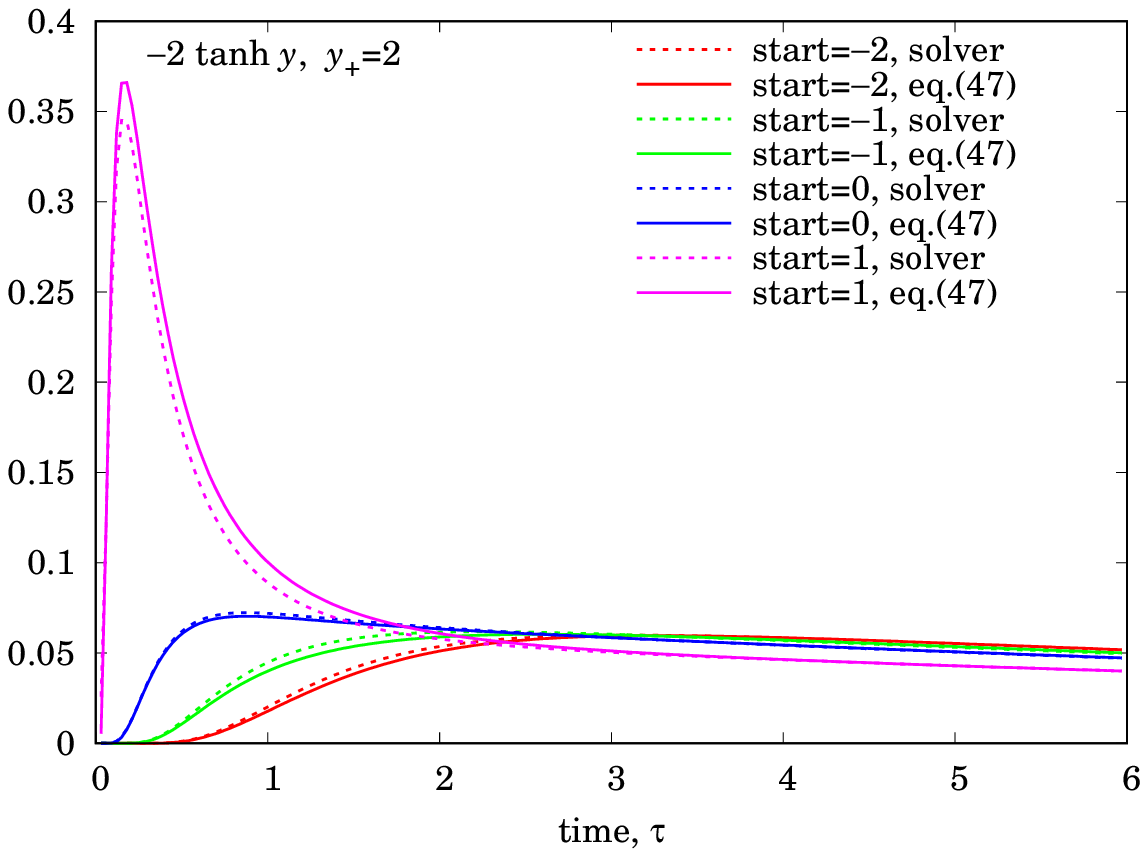}} \\
\end{tabular}
\caption{
First-passage time density for $\ffield(y)=-2\tanh y$: Eq.(\ref{eq:final}) compared with numerical PDE solver. Boundaries and starting-points as indicated on each plot.
}
\label{fig:sech2}
\end{figure}

\begin{figure}[!htbp]
\hspace{-10mm}
\begin{tabular}{llll}
(i) & \scalebox{0.6}{\includegraphics{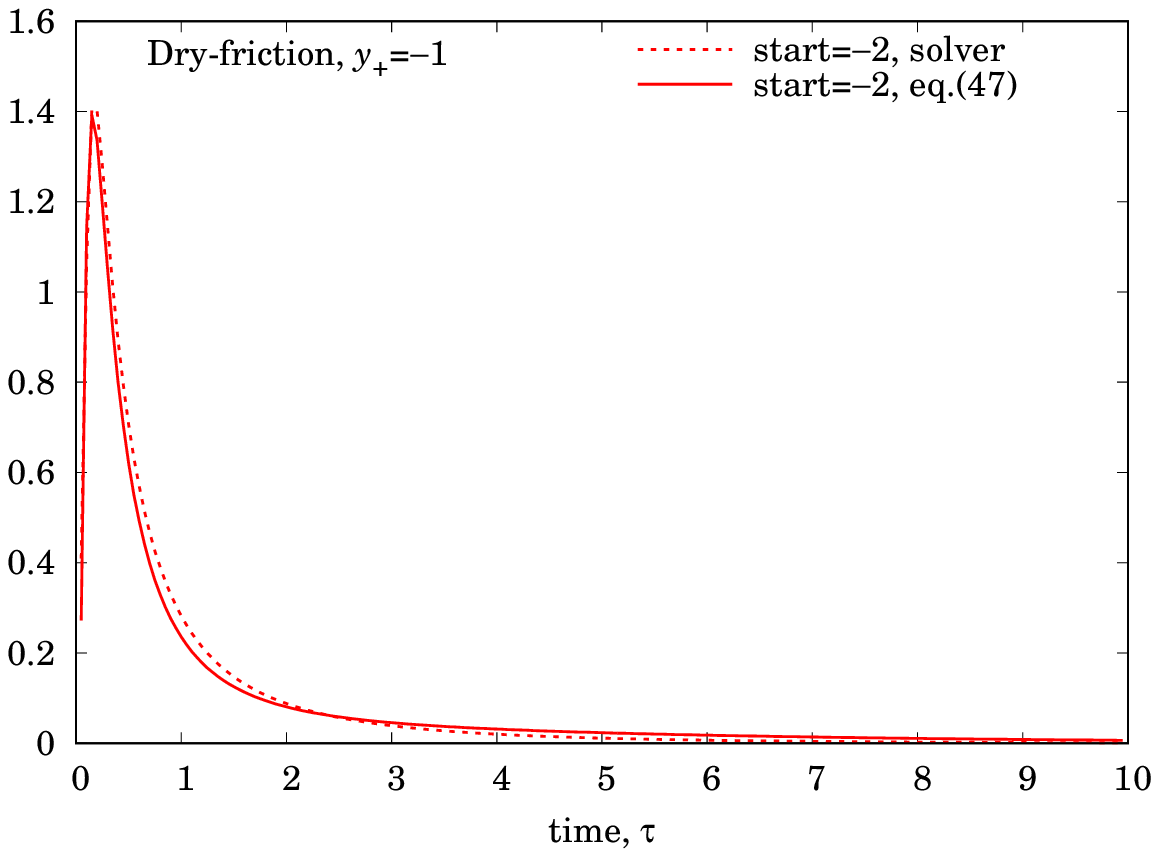}} &
(ii) & \scalebox{0.6}{\includegraphics{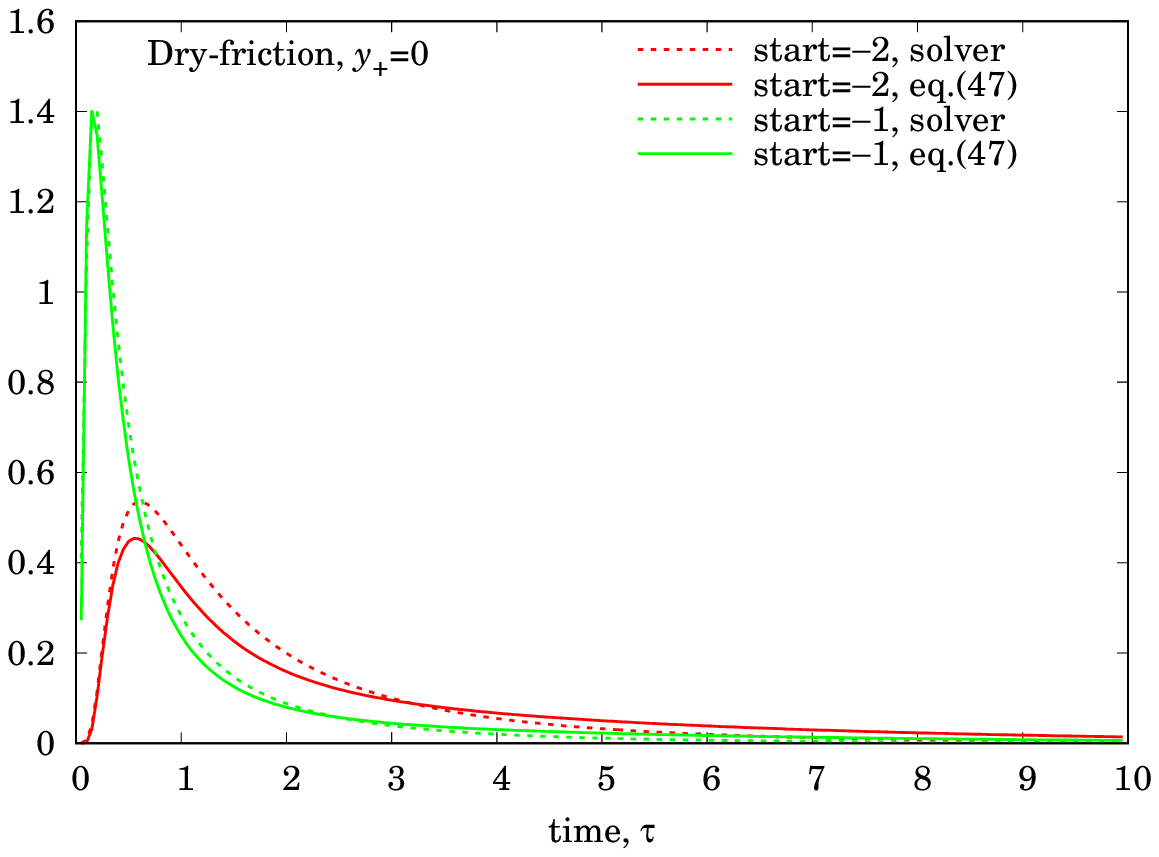}} \\
(iii) & \scalebox{0.6}{\includegraphics{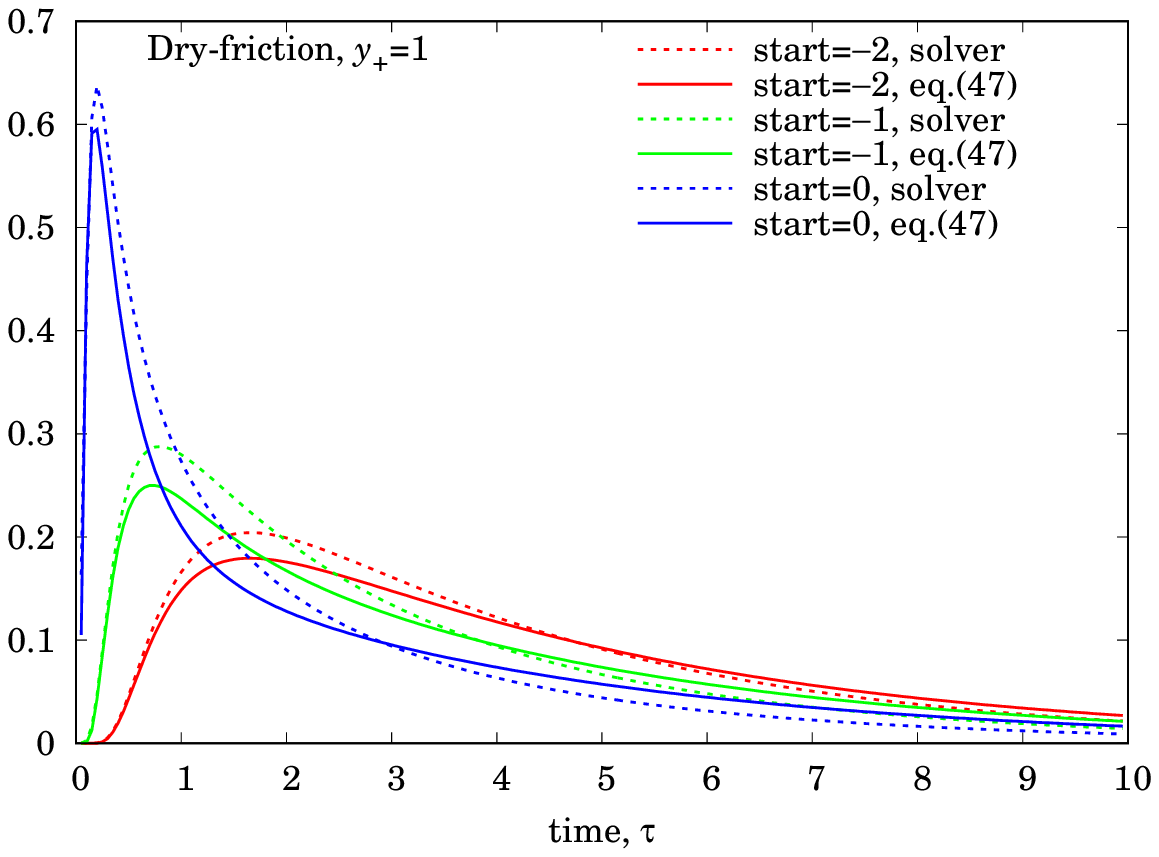}} &
(iv) & \scalebox{0.6}{\includegraphics{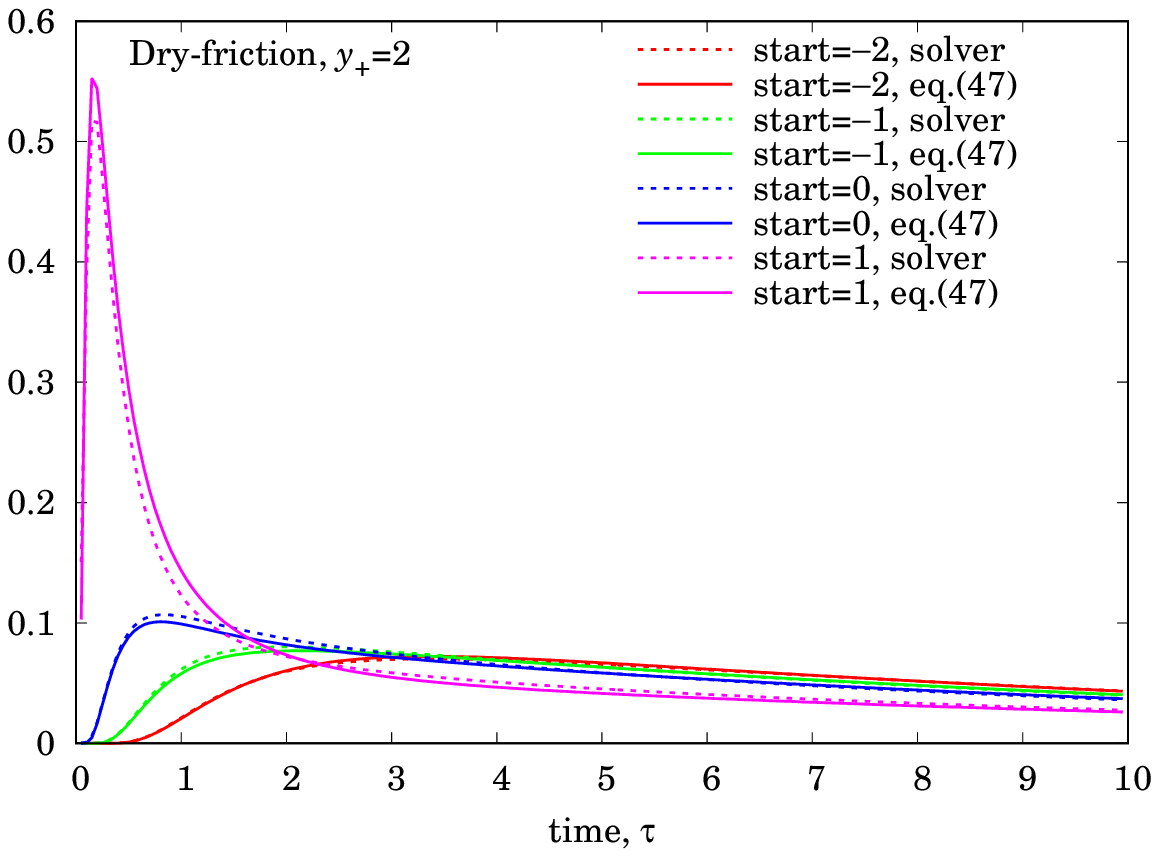}} \\
\end{tabular}
\caption{
First-passage time density for $\ffield(y)=-\sgn y$: Eq.(\ref{eq:final}) compared with numerical PDE solver. Boundaries and starting-points as indicated on each plot.
}
\label{fig:df}
\end{figure}

\subsubsection{Arithmetic Brownian motion}

As in \S~\ref{sec:lambda} we briefly mention this, despite the fact that $\psi$ is not normalisable. We have $\ffield(y)=\mu\ge0$, $\psi(y)=e^{\mu y}$, $\lambda=\mu^2/4$ and $\theta=0$ which we understand by taking the limit as $\theta\to0$ from above. The last two terms of (\ref{eq:final}) are unity. This converges to the inverse Gaussian distribution, as expected. Incidentally it also works when $\mu<0$, despite the fact that this case is not covered by the hypotheses of the paper: note that the first-passage time density no longer integrates to unity, as there is a positive probability of never hitting the boundary.

\subsubsection{$\ffield(y)=-2\tanh y$}

As before, we take $\ffield(y)=-2\tanh y$, giving $\psi(y)=\half \sech^2 y$.
% barr = -2, -1, 0, 1, 2, 3
% lambda = 1, 1, 1, 0.420, .0706, .00987
% theta = 4/3
Figure~\ref{fig:sech2} shows the results, and again the agreement is good.

\subsubsection{$\ffield(y)=-\sgn y$ (dry-friction case)}

As before, we take $\ffield(y)=-\sgn y$, giving $\psi(y)=\half e^{-|y|}$.
% barr = 1, 2
% lambda = .25, .0913
% theta = 1
In calculating (\ref{eq:ecoef}) we always take $\ffield'(\barrier)=0$, even when $\barrier=0$ (to be understood as the limit $\barrier\nearrow0$).
Figure~\ref{fig:df} shows the results. 
For this model the agreement is less good, particularly when $\barrier\le0$, though the short-term behaviour is correct and the long-term rate of exponential decay is correctly captured: as we said in \S\ref{sec:short}, it is $e^{-\tau/4}$ whenever $\barrier\le1$. 
What makes this model difficult to approximate is that it is essentially two very different models joined together.
If the boundary is below the equilibrium level, the model is just an arithmetic Brownian motion and there is no mean reversion\footnote{As, by convention in this paper, we have chosen to start below the boundary.}. If the boundary is above the equilibrium level, the model becomes mean-reverting and exhibits different behaviour. Either model can be successfully approximated on its own---indeed, as we said above, (\ref{eq:final}) is exact for the arithmetic Brownian motion.   However, we do not have the luxury of being able to take two different copies of (\ref{eq:final}), with different parameters, to represent the two halves, and the results shown are the consequence of trying to encapsulate all the properties of the model into one.

%%%%%%%%%%%%%%%%%%%%%%%%%%%%%%%%%%%%%%%%%%%%%%%%%%%%%%%%%%%%%
%%%%%%%%%%%%%%%%%%%%%%%%%%%%%%%%%%%%%%%%%%%%%%%%%%%%%%%%%%%%%

\section{Conclusions}

We have derived an approximate expression for the first-passage time density of a mean-reverting process, that captures the short- and long-term behaviour in a single formula, Eq.(\ref{eq:final}).
The development has used the Ornstein--Uhlenbeck process as a prototype, and in certain cases delivers exact results. However, it possesses much greater generality than that, and our basic thesis is that within a broad class of models the answer can always be effectively approximated this way.
Perhaps the most cogent reason for wanting to work with an expression resembling (\ref{eq:final}) is that it \emph{looks} like the stopping-time density of a mean-reverting diffusion: in other words it has a coherent form in a way that a Bromwich integral or an eigenfunction expansion does not.

The paper does not pretend to be the last word on the subject.
It is likely that the most productive approach to this problem is a combination of analytical and numerical techniques: the latter may include the numerical solution of integral equations, or the use of spectral methods \cite{Boyd01}.
In principle the formula (\ref{eq:final}), coupled with (\ref{eq:pde_Fy}), permits such an approach. If we extract the most important terms from (\ref{eq:final}), and write\footnote{We are re-using the symbol $R$.}
\[
f(\tau,y) = 
\frac{(\barrier-y)e^{-\lambda \tau}}{\sqrt{\pi(1-q)^3/2\theta^3}}  
\exp \left( \frac{- \theta\sqrt{q}(y-\barrier)^2}{2(1-q)} \right)
\left( \frac{\psi(y_+)}{\psi(y)} \right)^{ \textstyle \frac{\sqrt{q}}{1+\sqrt{q}} }  
\big( 1+ R(\tau,y)\big) ;
\]
then $R$ satisfies a parabolic PDE and, by construction, it is known to be zero in several different limits; as it is smooth and slowly-varying, it is an ideal candidate for approximation by spectral methods. 
%*** I'm not totally sure about this. I think it may be that in some limits you can only say that $\partial R/\partial y\to0$, or $\partial R/\partial \tau\to0$, which is weaker. This isn't a show-stopper, but needs toning down. Certainly $R$ is zero at short time ***
(Another possibility is to replace $1+R$ with $e^R$, which ensures positivity at the expense of creating a nonlinear PDE.)
This method of attack has been applied to special functions for many years: the basic idea is to extract various factors and/or transform the function in question by considering its behaviour in various limits, and then approximate the remainder term with a Chebyshev expansion or some variant of it. Many of the approximations in \cite{Abramowitz64} fall into this category.
Another possibility is to use the above definition and develop the remainder term numerically using the Volterra integral equation techniques of \cite{Lipton18}.

Possible further developments include multidimensional analogues (the exit time from a polygonal zone, for example), one-dimensional problems with two boundaries, commonly called the 
double-barrier problem or exit time from a channel, which have been studied in  e.g.~\cite{Dirkse75,Donofrio18,Lindenberg75,Sweet70}.
Another possibility is the first-passage time of a L\'evy process rather than simply a diffusion, for which recent discussions and applications are in e.g.~\cite{Lipton02c,Martin10b,Martin18a}. While the long-time behaviour in such models is still exponential, the short-time behaviour is typically different.

\section{Acknowledgements}

RM thanks Ridha Nasri for his advice on parabolic cylinder functions and Nicholson integrals, and Alexander Lipton for helpful insights into PDE theory. We are also grateful to Satya Majumdar for many discussions on first-passage time problems over the years.

%%%%%%%%%%%%%%%%%%%%%%%%%%%%%%%%%%%%%%%%%%%%%%%%%%%%%%%%%%%%%%
%%%%%%%%%%%%%%%%%%%%%%%%%%%%%%%%%%%%%%%%%%%%%%%%%%%%%%%%%%%%%%
%%%%%%%%%%%%%%%%%%%%%%%%%%%%%%%%%%%%%%%%%%%%%%%%%%%%%%%%%%%%%%
%\clearpage
\appendix

\section{Appendix}

\subsection{Reflection formula (\ref{eq:pcfrecur2},\ref{eq:pcfrecur3}) for $\bigD_s(z)$}

For the proof of (\ref{eq:pcfrecur2}), write the LHS as
\[
\frac{1}{\Gamma(s)\Gamma(1-s)}
\int_0^\infty u^{s-1} e^{yu-u^2/2} \, du
\int_0^\infty v^{-s} e^{yv-v^2/2} \, dv
\]
and change variables by $u=tw$, $v=(1-t)w$ to obtain
\[
\frac{1}{\Gamma(s)\Gamma(1-s)}
\int_0^\infty \int_0^1  t^{s-1}(1-t)^{-s} e^{yw-w^2/2} e^{w^2 t(1-t)} \, dt \, dw.
\]
Then expand the last exponential as a Taylor series: the $t$-integral can then be done using the Beta function and the $w$-integral is another parabolic cylinder function. The derivation is valid only for $\Real s\in (0,1)$, but the result extends to $s\in\C$ by analytic continuation. Then (\ref{eq:pcfrecur3}) is obtained by writing $\bigD_{2k+1}(y)$ in terms of its integral definition and then summing the series.
Note in passing the attractive result
\[
\big[ \bigD_\half(y) \big] ^2 =  \sum_{k=0}^\infty \frac{\Gamma(k+\half)^2}{\pi k!} \bigD_{2k+1}(y) = \int_0^\infty {}_1F_1(\shalf;1;z^2/4) \, e^{yz-z^2/2} \, dz  
 .
\]

\subsection{$\lambda$ vs boundary for OU case}

As the OU model is a popular one, it is worth tabulating the exact values of $\lambda$ as a function of $\barrier$ at selected points, which we have done in Table~\ref{tab:1}. For intermediate values we suggest  polynomial interpolation of $\lambda$ when $\barrier<0$, and polynomial interpolation of $\ln \lambda$ when $\barrier>0$. This allows a good approximation to be calculated very rapidly.

\begin{table}[!htbp]
\centering
\begin{tabular}{rr}
\hline
$\barrier$ & $\lambda$ \\
\hline
$-2.86$ & 5 \\
$-2.33$ & 4 \\
$-\sqrt{3}$ & 3 \\
$-1$ & 2 \\
$-0.5$ & 1.449 \\
0 & 1 \\
0.5 & 0.649   \\
1 & 0.388  \\
1.5 & 0.209  \\
2 & 0.0973  \\
2.5 & 0.0377  \\
3 & 0.0116  \\
\hline
\end{tabular}
\caption{$\lambda$ vs boundary position, for OU. (Except where results are exact they are given to 3 s.f.)}
\label{tab:1}
\end{table}

\subsection{Continuing the short-time development of $h$}

We turn to ideas that surround the extension of (\ref{eq:ansatz1}).
Let us consider the following representation for $h$, as obtained by earlier discussion:
\begin{equation}
h(\tau,y) = \frac{y-\barrier}{2\tau} + \frac{\ffield(y)}{2} + \frac{1}{\barrier-y} + R(\tau,y), \qquad \tau\ll 1,
\end{equation}
where $R$ is a residual term to be found and/or approximated.
Provided $R(\tau,y)\to0$ as $\tau\to0$ the short-time behaviour will be correct. Further, we require $R(\tau,\barrier)$=0 for all $\tau$ so that the behaviour on the boundary is correct too: more precisely, the error in $h$ will be $O(y-\barrier)$.

Substitution of the above expansion into (\ref{eq:pde_hy}) gives
\begin{equation}
\pderiv{R}{\tau} = \pdderiv{R}{y} + \pderiv{}{y} \left(  \frac{2R}{y-\barrier} + \frac{\barrier-y}{\tau}R  \right) 
+ \pderiv{}{y} \left( -R^2 + \frac{\ffield(y)^2}{4} + \frac{\ffield'(y)}{2}  \right),
\label{eq:R1}
\end{equation}
a parabolic PDE in which the last group of terms can be thought of as a forcing term.

Apparent from the PDE is that there is an interplay between $(y-\barrier)^2$ and $\tau$, suggesting a change of variable to $(\tau,v)$ defined by
\[
v = (y-\barrier)^2/2\tau;
\]
in any case this kind of similarity solution (distance proportional to square root of time) is obvious on intuitive grounds.
Then
\[
\left( \pderiv{}{\tau} \right)_y = \left(\pderiv{}{\tau}\right)_v - \frac{v}{\tau} \pderiv{}{v} , \qquad
\left( \pderiv{}{y} \right)_\tau = -\sqrt{\frac{2v}{\tau}} \pderiv{}{v}
\]
where $(\partial/\partial \tau)_v$ means keeping $v$ constant, and so on. So (\ref{eq:R1}) becomes
\[
\left(\pderiv{R}{\tau}\right)_v = \frac{2v}{\tau} \pdderiv{R}{v} + \frac{3}{\tau} \pderiv{R}{v} - \frac{v}{\tau} \pderiv{R}{v} - \frac{R}{\tau} - \frac{R}{v\tau} \\
+ \mbox{(Forcing)}
\]
which can also be written
\begin{equation}
\left(\pderiv{R}{\tau}\right)_v = \frac{1}{\tau v} \left( 2v \pdderiv{}{v} - (1+v) \pderiv{}{v} \right) v R 
+ \mbox{(Forcing)} .
\label{eq:R2}
\end{equation}

As $R$ is initially zero, and zero at the boundary, we can attempt to understand the short-time behaviour by ignoring the $R^2$ term. The resulting PDE is then linear and easily solved, at least in certain cases, and we denote by $\widetilde{R}$ the solution to the linearised equation.
The first thing to notice is that in the absence of forcing, separation of variables gives solutions of the form\footnote{We are abusing the notation by writing $\widetilde{R}(\tau,y)$ and also $\widetilde{R}(\tau,v)$ but this should not cause confusion.}
\[
v \widetilde{R}(\tau,v) = \tau^a U(a,-\shalf,v/2)
\]
where $U$ denotes the confluent hypergeometric function of the second kind \cite{Abramowitz64}\footnote{In other words the Tricomi function. The Kummer function grows exponentially as $v\to\infty$, which cannot give the right behaviour.},
\[
U(a,b,z) = \frac{1}{\Gamma(a)} \int_0^\infty x^{a-1}(1+x)^{b-a-1} e^{-zx} \, dx.
\]
Consider next a forcing term of the form $\tau^{\alpha} v^\beta$.
\notthis{
By trying a solution of the form
\[
\widetilde{R}(\tau,v) = k \tau^{1+\alpha} v^c U(a,b,v/4)
\]
and solving for $a,b,c,k$, we find
\[
\widetilde{R}(\tau,v) = 
\]
} % end cut
It is natural to seek a solution to (\ref{eq:R2}) of the form 
\[
v\widetilde{R}_{\alpha,\beta} (\tau,v) = \tau^{\alpha+1} g_{\alpha,\beta}(v)
\]
where $g_{\alpha,\beta}$ is a root of 
\[
2v g''(v) - (1+v) g'(v) - (1+\alpha) g(v) = -v^{1+\beta}.
\]
\notthis{
The complementary problem (zero RHS) is solved by the confluent hypergeometric function, necessarily of the second kind :
\[
U(1+\alpha, - \shalf, v/4) = \frac{1}{\Gamma(1+\alpha)} \int_0^\infty x^{\alpha}(1+x)^{-\alpha-5/2} e^{-vx/4} \, dx.
\]
The full solution is then obtained by taking a particular integral plus some multiple of this, so as to obey the boundary condition $R=0$ at $v=0$.
} % end notthis
As examples:
\begin{itemize}
\item
$\alpha=\beta=0$. A particular solution is $(v-1)/2$, giving
\[
g_{0,0}(\tau,v) =  \frac{v-1}{2} + \frac{3}{4} U(1,-\shalf,v/2) = \frac{v^2}{4} U(1,{\textstyle\frac{3}{2}},v/2) ,
\]
the latter following from repeated integration by parts. Noting that
\[
U(1,{\textstyle\frac{3}{2}},z) \equiv \sqrt{\pi/z} \, e^z \erfc\sqrt{z}
\]
we can recast the solution in the original coordinates as
\[
\widetilde{R}_{0,0} (\tau,y) = \frac{\tau}{2} \frac{\gothicz  \Phi(-\gothicz)}{\phi(\gothicz)} , \qquad \gothicz = \frac{\barrier-y}{\sqrt{2\tau}}.
\]
The solution grows as $\tau/2$ in the `outer' zone $\gothicz\gg1$; the other part of the expression, containing the probability integral, expresses the behaviour at and near the boundary. 
\item
$\alpha=\beta=\half$. This case may be solved at once to give
\[
\widetilde{R}_{\half,\half}(\tau,y) = \frac{\tau(\barrier-y)}{3}.
\]
It occurs in the OU model when $\barrier=0$, as in effect we are expanding $\frac{q}{1-q}$ as a Laurent series in $\tau$, invoking the Bernoulli numbers.

\end{itemize}

The OU case is now analysed as follows:
\[
\pderiv{}{y} \left( \frac{\ffield(y)^2}{4} + \frac{\ffield'(y)}{2} \right) = \frac{y}{2} = \frac{\barrier}{2} - \sqrt{\frac{\tau v}{2}}
\]
and so the linearised solution is
\begin{equation}
\widetilde{R}(\tau,y) = \frac{\barrier}{2} \widetilde{R}_{0,0}(\tau,y) - \frac{1}{2} \widetilde{R}_{\half,\half}(\tau,y)
 = \frac{\tau \barrier}{4} \frac{\gothicz \Phi(-\gothicz)}{\phi(\gothicz)} - \frac{\tau(\barrier-y)}{6},
\end{equation}
with $\gothicz=(\barrier-y)\big/\!\sqrt{2\tau}$ as stated above.

This presents a different conclusion from that of \cite{Martin15b}, wherein the expansion for the remainder term was
\[
R(\tau,y) = q\sum_{r=1}^\infty (1-q)^r b_r(y).
\]
Such an expansion does not work here: one cannot find $b_r(y)$ that are regular and obey $b_r(\barrier)=0$. We have chosen not to go into the details of this, preferring instead to say what the solution \emph{does} look like. The presence of terms that are essentially singular when $\gothicz\to\infty$ makes it clear that the solution is not analytic at $\tau=0$; more precisely a Taylor series in $\tau$, with coefficients that are functions of $y$, will not work. An expansion in powers of $\tau$ with coefficients that are functions of $\gothicz$ is not precluded, however.

\bibliographystyle{plain}
%\bibliography{phd}

\end{document}